\documentstyle[pre,aps,multicol,epsf]{revtex}

\def\marginpar#1{ }
\def\eqn#1#2{\begin{equation}#1
             \label{#2}
              \end{equation}
}
\def\h{\hbar}
\def\pd{\partial}
\def\di{{\rm d}}

\def\bj{{\bf j}}
\def\bFl{{\bf f}}
\def\bk{{\bf k}}
\def\bl{{\bf l}}
\def\bp{{\bf p}}

\def\br{{\bf r}}
\def\bs{{\bf s}}
\def\bx{{\bf x}}
\def\bv{{\bf v}}
\def\by{{\bf y}}
\def\bz{{\bf z}}
\def\bA{{\bf A}}
\def\bB{{\bf B}}

\def\bE{{\bf E}}

\def\bG{{\bf G}}
\def\bH{{\bf H}}

\def\bP{{\bf P}}
\def\bR{{\bf R}}
\def\bX{{\bf X}}
\def\bpi{\mbox{\boldmath$\pi$}}
\def\bSigma{{\mbox{\boldmath$\Sigma$}}}
\def\bnabla{\mbox{\boldmath$\nabla$}}

\def\bomega{\mbox{\boldmath$\omega$}}
\def\deg{{\mbox{\dag}}}

\def\dbarPn{\overline{\overline{P^n}}}
\def\dbarbPb{\overline{\overline{{\bf P}^2}}}
\def\dbarvarep{\overline{\overline{\varepsilon}}}
\def\dbarP{\overline{\overline{P}}}
\def\dbarbP{\overline{\overline{{\bf P}}}}

\def\dbarPPl{\overline{\overline{\bP_\lambda\bP}}}
\def\dbarv{\overline{\overline{{\bf v}}}}
\def\dbarJ{\overline{\overline{{\bf J}}}}
\def\e{{\rm e}}
\def\dint{\displaystyle\int\limits}
\def\slashnabla{\nabla
                \hspace{-.7em}\mbox{\raisebox{+.2ex}{$/$}}
            \hspace{+.2em}}
\def\slashbnabla{\nabla
                \hspace{-.7em}\mbox{\raisebox{+.2ex}{$/$}}
            \hspace{+.2em}}
\def\slashpartial{\partial
                \hspace{-.5em}\mbox{\raisebox{+.2ex}{$/$}}}
\def\slashvarepsilon{\varepsilon
                \hspace{-.5em}\mbox{\raisebox{+.2ex}{$/$}}}
\def\slashbP{{\bf P}
                \hspace{-.7em}\mbox{\raisebox{+.2ex}{$/$}}
            \hspace{+.2em}}
\def\slashbA{{\bf A}
                \hspace{-.7em}\mbox{\raisebox{+.2ex}{$/$}}
            \hspace{+.2em}}

\def\etal{{\em et al}}
\def\wqdf{{Wigner quasi-distribution func\-tion}}
\def\rightlimit#1{
                \mbox{$
                        {\raisebox{-.34em}{$\overline{\scriptstyle{#1}
                                \mbox{{$\scriptstyle{A}$}}
                                                      }
                                           $}
                        }
                     $}
               \hspace{-.7em}\mbox{$\rightarrow$}
                }

\begin{document}

%






\title{Wigner quasi-distribution function
for charged particles in classical electromagnetic fields}
\author{M Levanda\footnote{email: matty@levanda.co.il}
 and V Fleurov\footnote{email: fleurov@post.tau.ac.il }\\
\author{V Fleurov}
Raymond and Beverly Sackler Faculty of Exact Sciences, \\ School
of Physics and Astronomy, Tel Aviv University, \\ Tel Aviv
69978 Israel } 

\maketitle






\begin{abstract}A gauge-invariant \wqdf\ for charged particles in classical
electromagnetic fields is derived in a rigorous way. Its relation
to the axial gauge is discussed, as well as the relation between
the kinetic and canonical momenta in the Wigner representation.
Gauge-invariant quantum analogs of Hamilton-Jacobi and Boltzmann
kinetic equations are formulated for arbitrary classical
electromagnetic fields in terms of the `slashed' derivatives and
momenta, introduced for this purpose. The kinetic meaning of these
slashed quantities is discussed. We introduce gauge-invariant
conditional moments and use them to derive a kinetic momentum
continuity equation. This equation provides us with a hydrodynamic
representation for quantum transport processes and a definition of
the `collision force'. The hydrodynamic equation is applied for
the rotation part of the electron motion. The theory is
illustrated by its application in three examples. These are:
\wqdf\ and equations for an electron in a magnetic field and
harmonic potential; \wqdf\ for a charged particle in periodic
systems using the $kq$ representation; two \wqdf s for heavy-mass
polaron in an electric field. \end{abstract}


\vspace{1cm}



%
\section{Introduction \label{intwig}}
In 1932 Wigner introduced a possible method for describing a
quantum system in terms of a phase space quasi-distribution
function which is now known as the \wqdf. It allows quantum
mechanical expectation values to be written in the form of phase
space integrals similarly to the corresponding expectation values
in classical mechanics (see, e.g., Balazs and Jennings, 1984;
Hillery \etal , 1984). The \wqdf\ is closely connected with the
Weyl (1931) correspondence principle (see also Schwinger, 1960),
which has been originally formulated for systems free of external
electromagnetic fields.

The \wqdf,
$$\rho_W(P_{can},X) = \rho_W( \varepsilon_{can},
\bP_{can},T,\bR),$$
is defined as an expectation value of the so-called Wigner
operator, $\hat\rho_W(P_{can},X)$ obtained by a coordinate
transformation of a single particle density operator,
$\hat\rho(x,x')= \hat\rho(t,\br,t',\br')$. This transformation
Wigner (1932) was chosen in the form
\eqn{
\hat\rho_{W}(P_{can},X)= \dint \di^4 y\e^{-(i/\h) P_{can}^\mu y_\mu}
\hat\rho(y,X)}
{Fourier1}
where $X=(x+x')/2$, $y=(x'-x)$ and the subscript $can$ indicates
that $P$ in this expression is the variable conjugated to the
coordinate $X$. Although the transformation (\ref{Fourier1}) seems
quite obvious, it has a highly significant drawback of producing
not gauge-invariant quantities for charged particles in an
electromagnetic field. The resulting conjugate variable $P_{can} =
(\varepsilon_{can}, \bP_{can})$ does not necessarily have a
meaning of the kinetic four momentum.

Here lies an important difference between the classical and
quantum descriptions. In classical mechanics we deal with the
well-defined coordinate and momentum of the particle. The
electromagnetic field is introduced in the Hamilton function of
the system simply by considering the canonical momentum
$$P_{can}=P_{kin}+{e\over c}A(x)$$
instead of the kinetic momentum. Here $A^\mu(x) = (\varphi(x),\
\bA(x))$ is the four vector-potential of the applied
electromagnetic field. While it is well known that no problems
with defining gauge-invariant quantities arise in classical
mechanics, the uncertainty principle does not allow one to define
simultaneously a coordinate and a momentum of a quantum particle.
Quantum mechanics must, therefore, deal with a two-point density
operator $\hat\rho(x,x')= \hat\rho(t,\br,t',\br')$. Then a gauge
transformation converts the potential $A(x)$ differently in these
two different points, making an introduction of the canonical
momentum not so harmless and even ambiguous procedure. The result
is that equation (\ref{Fourier1}) is not the only, nor is it the
best definition of the \wqdf\ for a system in an electromagnetic
field.

Nonetheless, the transformation (\ref{Fourier1}) has been used
many times in the past for systems in electromagnetic fields. This
may lead to certain difficulties\footnote{Correct calculations
using noninvariant quantities lead, nevertheless, to correct gauge
invariant physical quantites (expectation values). The above
mentioned references and section \ref{examples} contain examples
of misinterpretations, approximations and assumptions that may
result from attempts to provide physical reasoning to
non-gauge-invariant equations, expressions and other terms used
during the calculations. The discussion of the meaning of the
slashed derivatives presented in subsection \ref{slashsec} implies
that one should be cautious when considering the Poisson brackets
for a non-gauge-invariant Boltzmann kinetic equation.} in
calculations and in interpretation of the results (see, e.g.,
Mahan, 1987; Mahan, 1990; Rammer, 1986, and the discussion in
Levanda and Fleurov, 1994) since terms explicitly depending on the
potential may appear, e.g., in kinetic equations.

Fleurov and Kozlov (1978) (see also Al'tshuler, 1978) applied a
transformation of the form
\eqn{
\hat\rho_W(P,X)=
\dint\di^4y \e^{-{i\over\h}y^\mu (P_\mu -{e\over c}A_\mu(X))}
\hat\rho(y,X),}
{Fourier2}
which produces explicitly gauge-invariant quantities and equations
in the linear approximation or in a nonlinear case of a static
homogeneous electric field (Tugushev and Fleurov, 1983; Serimaa
\etal , 1986 ).

The first who tried formulating a gauge-invariant \wqdf\ for an
electron in classical electromagnetic fields was Irving (1965) who
proposed to employ the transformation
$$\hat\rho_W(P,X)=$$
\eqn{\dint \di^4 y \exp\left( -{i\over\h} P y + {i e\over c\h}
\dint^{1/2}_{-1/2} y^\mu A_\mu
 (X+sy)\di s
 \right) \hat\rho(y,X)}
{Fourier3}
which is truly gauge-invariant. It is worthwhile to pay a
particular attention to the phase factor in (\ref{Fourier3})
emerging from the proper time method of Schwinger (1951). This
phase factor is related to the dynamics of the charged particle in
the electromagnetic field. Really, the phase factor of an electron
wavefunction which transforms as
\eqn{
\psi (x)\rightarrow\psi(x)\e^{(ie/\h c)\Lambda (x)}}
{gauge1}
under the local gauge transformation
\eqn{
A_\mu (x)\rightarrow A_\mu (x)-\pd_\mu\Lambda(x)}
{gauge2}
with an arbitrary real function $\Lambda (x)$, compensate for the
change in the phase factor in equation (\ref{Fourier3}), and the
Wigner operator $\hat\rho_W(P,X)$ remains unchanged. This ensures
the gauge invariance of the definition (\ref{Fourier3}).

It is, however, quite clear that the choice of the integration
path is still ambiguous and is not anyhow restricted by the
condition of the gauge invariance. Another path of integration
differing from the straight line would mean a differing definition
of the variable $P$. This is the phase factor
\eqn{
\exp\left({i e\over c\h} \dint^{X+y/2}_{X-y/2}
A_\mu(z)\di z^\mu \right)}
{line}
which would ensure the gauge invariance of the definition
(\ref{Fourier3}) for any integration path (Mandelstam, 1960).
Vasak \etal\ (1987) and Elze \etal\ (1986) proposed removing the
above ambiguity by choosing the straight integration path. They
showed that for this choice  the transform (\ref{Fourier3})
becomes equivalent to the transform (\ref{Fourier1}) when
substituting the covariant derivative $\pd^X+(e/c)A(X)$ for
$\pd^X$.

In spite of all that has been done, the situation remains rather
confusing. The phase factor is introduced in equation
(\ref{Fourier3}) only as a means of solving the problem of the
gauge invariance. The arguments in favor of a specific phase
factor are mainly intuitive. Hence, it is not quite clear which
transformation is to be used if one prefers choosing other
variables in the \wqdf\ or if the dispersion relations for the
charged particles differ from quadratic. It is our aim in this
paper to derive systematically the transformation (\ref{Fourier3})
and to illustrate the importance and physical significance of the
phase factor by several examples.

Subsection \ref{giwqdf} formally defines the variables of the Weyl
correspondence principle to be the kinetic momenta and the
coordinates. We show that the form of the transformation
(\ref{Fourier3}) is due to the minimal coupling principle used to
relate the canonical and kinetic momenta. The phase factor
introduced in (\ref{Fourier3}) is shown to be a gauge
transformation that transforms an arbitrary gauge into an axial
gauge. Working with this axial gauge, the vector potential
vanishes along the integration trajectory. In addition the kinetic
and canonical momenta coincide and the two transforms,
(\ref{Fourier1}) and (\ref{Fourier3}), are identical. We shall
dwell on the connection between the kinetic and canonical momentum
in the Wigner representation as well as on its classical limit.

Quantum kinetic processes are in many cases treated by means of
the Kubo linear response method Kubo (1957). An alternative
approach, based on Keldysh (1965) diagrammatic technique, enables
one to go rather easily beyond the linear response approximation.
The first to formulate kinetic equations, based on the real time
Green's function technique, introduced by Martin and Schwinger
(1959), were Kadanoff and Baym (1962). Fleurov and Kozlov (1978)
using Keldysh formulation deduced two sets of linearized quantum
kinetic equations which may in principle give an exhaustive
description of the dynamic and kinetic properties of a system of
interacting Fermi-particles. Since then a number of papers were
published on quantum kinetic equations (Mahan, 1987; Rammer and
Smith, 1986; Tugushev and Fleurov, 1983; Reizer and Sergeev, 1987;
Khan \etal , 1987; Reggiani \etal , 1987; Davies and Wilkins,
1988; Bertoncini \etal , 1989; Ferry and Grubin, 1995) and
references therein. A gauge invariant formulation of the quantum
kinetic equations for charged particles in an arbitrary classical
electromagnetic field was derived in our paper (Levanda and
Fleurov, 1994). Here, Section \ref{kineticequations}, briefly
revisits this derivation and represents the resulting quantum {\em
Hamilton-Jacobi}\ and {\em Boltzmann}\ equations in a more compact
form by using the `slashed' derivatives. These derivatives make
the kinetic meaning of various terms in the equations more
transparent.

A hydrodynamic representation of the quantum theory was introduced
by Madelung (1927) who transformed the Schr\"{o}dinger equation,
for a free particle, into particle and momentum continuity
equations in the form similar to those found in hydrodynamic
theories. We use our quantum kinetic equations in the same manner
to obtain a hydrodynamic representation for description of
dynamics of {\em interacting} charged particles in arbitrary
external electromagnetic fields. The term hydrodynamic in this
context may be somewhat misleading, as the reader may infer that
`hydrodynamic' approximations are implied. This, however, is not
the case. The equations to be obtained would not assume any sort
of approximations. The full set of equations for conditional
moments (hydrodynamic equations) is equivalent to the original
Dyson equations and, hence, they  may account for all quantum
effects. It is emphasized that the hydrodynamic equations in
principle allow us to find the state of the system in and out of
equilibrium without further assumptions (neglecting interactions
these equations are equivalent to a single particle Schr\"odinger
equation). This is not the case with other transport equations
(e.g., for conventional quasi-classical Boltzmann equation). We
show that collisions lead to an introduction of a `collision
force' in the momentum continuity equation. The collision force
has the structure proposed by Bohm (1952) in his hidden variable
interpretation, to modify the Schr\"{o}dinger equation.

Section \ref{examples} presents a few examples that emphasize
different aspects of the \wqdf\ and its equations. The
Schr\"{o}dinger equation and its solutions for an electron in
harmonic potential and magnetic induction are found, in Subsection
\ref{Landau}, from the hydrodynamic equation. This reveals the
nature of the `velocity' in such a system. We place a particular
emphasis on the growth of the `velocity' with the coordinates. The
\wqdf\ for such a system is found from the quantum kinetic
equations. After integrating over the kinetic energy variable and
taking one of the fields to vanish we reduce our result to
expressions found in the literature for the \wqdf\ in a magnetic
induction or in a harmonic potential. The difference between the
gauge-invariant and non-gauge-invariant functions is discussed.

Subsection \ref{KQ} formulates the \wqdf\ for systems with
discrete translational symmetries both in time and/or space (an
example is an electron in a crystal). The acceleration theorem for
an electron in a crystal is discussed and its relation to the
gauge is pointed out. The Green's function of the heavy mass
polaron in an electric field is found from the kinetic energy
continuity equation. If the electric field vanishes, the Green's
function is the same as the one found in the literature. If an
electric field is present we find two solutions, one corresponding
to a polaron with constant kinetic energy and the other one
corresponding to a polaron with a diverging canonical energy. This
serves as an example for the kinetic meaning of the slashed
derivatives.

\section{Wigner quasi-distribution function
for a charged particle in electromagnetic fields}
An expression for a \wqdf\ for a charged particle in an external
electromagnetic field is derived. The coordinate transformation to
the Wigner coordinates (\ref{Fourier3}) is deduced rigorously from
the Weyl correspondence principle.
\subsection{Derivation of the \wqdf}
Weyl's correspondence principle suggests that an operator $\hat
C$, corresponding to a classical dynamical variable $C(X,P_{can})$
is obtained by replacing the variables $P_{can}$ and $X$ in the
Fourier transform
\eqn{
C(y,\varpi)= {1\over (2\pi\hbar)^4}\int\di P_{can}\di
X\e^{-(i/\h)(yP_{can}+\varpi X)} C(P_{can},X)}
{Weyl2}
by the Hermitian operators, $\hat P_{can}$ and $\hat X$,
representing these variables in quantum mechanics,
\eqn{
\hat C=
{1\over (2\pi\hbar)^4}\int\di y
\di\varpi\e^{(i/\h)(y\hat P_{can}+\varpi\hat X)}
C(y,\varpi).}
{Weyl}
The expectation value of the operator (\ref{Weyl}) reads
\eqn{
\left\langle \hat C \right\rangle =
\int\di X\left[\hat\rho(y',X)\hat C\right]_{y'\rightarrow 0}}
{expc1}
where the angular brackets indicate an ensemble average. The second
quantization allows one to present the density operator as
\eqn{
\hat\rho(y',X)=\Psi^\deg(X+y'/2)\Psi(X-y'/2).}
{den}
by means of the creation $\Psi^\deg(X)$ and annihilation $\Psi(X)$
operators. Then its expectation value becomes
\begin{eqnarray}
\lefteqn{ \left\langle \hat C \right\rangle = \int \di
X'\left[\left\langle \Psi^\deg(X'+y'/2)\hat C \Psi(X'-y'/2)
\right\rangle\right]_{y'\rightarrow 0} \nonumber } \\ &=& {1\over
(2\pi\hbar)^8}\int \di P_{can} \di X C(P_{can},X) \int \di
y\di\varpi \e^{-(i/\h)({y {P_{can}}+\varpi {X}})} \times \nonumber
\\ & & \int\di X'\left\langle\Psi^\deg(X')\e^{(i/\h)(y\hat
P_{can}+\varpi\hat X)} \Psi(X')\right\rangle \label{expc2}
\end{eqnarray}
\marginpar{expc2}
If the variable $P_{can}$ is chosen to be the canonical momentum,
one can follow Moyal (1949) and write
\begin{eqnarray}
\left\langle \hat C \right\rangle & = & {1\over (2\pi\hbar)^8}\int
\di P_{can} \di X C(P_{can},X) \int \di y\di\varpi
\e^{-(i/\h)(yP_{can}+\varpi X)} \times \nonumber \\ & & \int\di
X'\left\langle \Psi^\deg(X') \e^{iy\hat P_{can}/2\h}
\e^{i\varpi\hat X/\h} \e^{iy {\hat
P_{can}}/2\h}\Psi(X')\right\rangle \nonumber \\ & =& {1\over
(2\pi\hbar)^4}\int \di P_{can} \di X C(P_{can},X)
\rho_{Wcan}(P_{can},X). \label{expc3}
\end{eqnarray}
\marginpar{expc3}
\noindent where
$$\rho_{W}(P_{can},X)= $$
\eqn{\langle\hat\rho_{W}(P_{can},X)\rangle = \int \di y
\e^{-(i/\h)y P_{can}} \left\langle \Psi^\deg(X+y/2)\Psi(X-y/2)
\right\rangle} {wqdf1}
is the \wqdf. Properties and usage of it were discussed in the
introduction.

Equation (\ref{wqdf1}) defines the \wqdf, $\rho_W(P_{can},X)$
which has, however, a serious drawback. Although, the physical
quantity $\langle\hat C\rangle$ is certainly gauge-invariant, the
function $\rho_W(P_{can},X)$, taken separately, depends on the
gauge of the electromagnetic field. However, we are interested in
a gauge invariant definition of the \wqdf\ which can be derived by
doing some modifications in the current derivation.

\subsection{The need for a gauge-invariant \wqdf}
A question which we find necessary to address is why a
gauge-invarian \wqdf\ is really needed? In the papers referred to
in the Introduction a gauge invariant formulation was proposed
only as a convenient representation, meant either to help the
intuition or to directly correspond to a classical distribution
function, with the coordinates and the kinetic momenta as
variables in the classical limit of the \wqdf. We want to
emphasize that this is not the whole story, a non-gauge-invariant
\wqdf\ does not always allow one to calculate expectation values
of kinetic quantities. In particular we shall see that it is
impossible to calculate the expectation value of the kinetic
energy of a system in an electromagnetic field from a
non-gauge-invariant function as phase-space integrals.

For the Weyl correspondence principal to apply, the operator $\hat
C$ must be a function of a linear combination of the operators
$\hat X$ and $\hat P_{can}$ (Moyal, 1949). Although, $\hat X$ and
$\hat P_{can}$ can be any complementary operators, in the
following, we shall only consider the case where the two operators
are the coordinate operator and the momentum operator (kinetic or
canonical). The need for a linear combination of the operators can
be seen by substituting $C(aX + bP_{can})$ for $C(P_{can},X)$,
with $a,\ b$ being constants, in equation (\ref{Weyl2}), changing
the variables to $\xi = aX + bP_{can},$ $\eta = aX - bP_{can}$,
$\gamma = \varpi/2a + y/2b$,  $\mu = \varpi/2a - y/2b$.
$$\hat C=\int\di \mu \di \gamma \exp \left[{i\over\hbar}
\left((\gamma-\eta)b\hat P_{can}+(\gamma+\eta)a\hat X
\right)\right]\times$$
\eqn{ \int\di \xi \di \eta
\exp\left[-{i\over\hbar}(\gamma\xi+\mu\eta)\right] C(\xi).}
{Weyl4}
Integrating over the new variables, $\eta$, $\mu$, $\gamma$ and
then over $\xi$, one finds that $\hat C = C(a\hat X + b\hat
P_{can})$. The correspondence, $C= C(X,P_{can})$ goes into $\hat C
= C(\hat X,\hat P_{can})$, which exists only for linear
combinations of the operators.

We conclude that the operators $\hat X$ and $\hat P_{can}$ cannot
be chosen arbitrarily and the \wqdf\ should be set up for two
non-commuting operators chosen according to the measured quantity
$\left\langle \hat C \right\rangle$. In general a kinetic
operator, e.g., kinetic energy $P^2/2m$, not always can be written
as a linear combination of canonical momenta and coordinates,
hence, its expectation value not necessarily can be calculated
with a non-gauge-invariant \wqdf.

There are two cases in which equation (\ref{expc3}) can be
employed. The first is that of a system in homogeneous and static
electric and magnetic fields. Then the vector potential is a
linear function of the coordinates. Hence, the kinetic momentum is
a linear combination of the canonical momentum and the coordinate.
The second case is when one is interested in the expectation value
of the current (it is proportional to the momentum). This can be
expressed as the sum of the expectation values of the canonical
momentum and of the vector potential, with each of these
expectation values being calculated separately.

The gauge-invariant \wqdf\ will be formulated in terms of the
kinetic momentum and will provide us a possibility of calculating
expectation values of kinetic operators.

\subsection{Derivation of a gauge-invariant Wigner
quasi-distribution function\label{giwqdf}}

The Cartesian coordinates and the canonical momenta are not the
only possible basis operators (Scully and Cohen, 1987; Pimpale and
Razavy, 1988). We may choose $P$ to be the kinetic momentum then,
using the equality (\ref{eqll1}) derived in the Appendix A,
equation (\ref{expc2}) becomes
\begin{eqnarray}
\left\langle \hat C \right\rangle
& = & {1\over (2\pi\hbar)^4}\int \di P \di X C(P,X)
{1\over (2\pi\hbar)^4}\int \di y\di\varpi
\e^{-(i/\h)({y {P}+\varpi {X}})}
\times \nonumber \\ & &
\int\di X'\left\langle
 \left(\Psi^\deg(X')\e^{y{\pd\over\pd X}}\right)
\exp\left({i e\over c\h} \dint^1_0 y^\mu A_\mu
 (\hat X+s\hat y)\di s\right)
\Psi(X')\right\rangle
\nonumber \\
&=& {1\over (2\pi\hbar)^4}
\int \di P \di X C(P,X)
\rho_W(P,X).
\label{expc4}
\end{eqnarray}
\marginpar{expc4}
where
$$\rho_W(P,X)=\int \di y\e^{-(i/\h)y P}\times$$
\eqn{\exp\left({i e\over c\h} \dint^{1/2}_{-1/2} y^\mu A_\mu
 (X+sy)\di s\right)\left\langle
\Psi^\deg(X+y/2)\Psi(X-y/2)
\right\rangle}
{wqdf2}
Here, as opposed to equation (\ref{wqdf1}), the resulting \wqdf,
$\rho_W(P,X)$, incorporates a phase factor and coincides with the
definition (\ref{Fourier3}) postulated by Irving (1965).

It is appropriate to discuss here the connection of the straight
integration path with the axial gauge (see, e.g., Rohrlich, 1965).
A vector potential $A_\lambda(x)$ in any gauge can be transformed
to the axial gauge  by means of the transformation
\begin{eqnarray}
A_\lambda(x) &\rightarrow & A'_\lambda(x;x'') = \nonumber
\\&&
A_\lambda(x)+ \pd^x_\lambda \left( (x''-x)^\mu \dint^{1/2}_{-1/2}
A_\mu \left((x''+x)/2+s(x''-x)\right)\di s \right) = \nonumber
\\& &
(x''-x)^\mu\dint^{1/2}_{-1/2} \di s ({1\over2}-s) F_{\lambda\mu}
\left((x''+x)/2+s(x''-x)\right) =\nonumber
\\& &
\left.(x''-x)^\mu\frac{1-e^{-a}(1+a)}{a^2}
\right|_{a=(x''-x)^\nu\partial^\nu} F_{\lambda\mu}(x'') =
\nonumber
\\& &
 (x''-x)^\mu \left({1\over 2} - {1\over 6}(x''-x)^\nu\partial^\nu
+ \cdots\right) F_{\lambda\mu}(x''). \label{int22}
\end{eqnarray}
\marginpar{int22}
The resulting vector potential depends only on the electromagnetic
tensor
$$F_{\lambda\mu}(X)=\pd_\lambda A_\mu(X)-\pd_\mu A_\lambda(X)$$
and on the coordinate $x''$ which can be chosen arbitrarily. This
form of the vector potential has two important properties. In
regions free of currents and charges, $j_\lambda =(c\rho,\bj)$
generating the electromagnetic field,
$$\pd^\lambda A'_\lambda(x)=$$
\eqn{(x''-x)^\lambda\dint^{1/2}_{-1/2} \di s({1\over2}-s)^2
j_\lambda \left((x''+x)/2+s(x''-x)\right) =0} {div}
and
\eqn{
(x''-x)^\lambda A'_\lambda(x;x'')= 0.}
{axial}
Hence, this transformation converts any vector potential into a
transverse vector field.

Now substituting this axial vector potential in the definition
(\ref{wqdf2}) and taking $x''-x=a(x'-x)/2 = ay/2$ with $a$ being a
constant, one sees that the integral in the phase factor becomes
zero making the phase factor equal one. Therefore, the
gauge-invariant \wqdf\ coincides for this specific case with the
gauge non-invariant definition (\ref{wqdf1}).

All that has been done thus far can be simply restated as follows.
The choice of the axial gauge with $x''=ay+x$, ensures that the
vector potential indeed vanishes (see equation (\ref{axial}))
along the line that goes through $x''$ and $x$. The displacement
operator along this line, can equally be written as
$\exp\left(-iP_{can}^\mu (x''-x)_\mu\right)$ or as
$\exp\left(-iP_{kin}^\mu (x''-x)_\mu\right)$ and the \wqdf\ of the
canonical momentum variable coincides with the \wqdf\ of the
kinetic momentum variable. The function of the phase factor is to
`transform' an arbitrary gauge chosen for the vector potential
into the axial gauge.

Two remarks are in order. This derivation of the gauge-invariant
\wqdf, can be applied to other correspondence principles as well,
which in turn will lead to other quasi-distribution functions. As
was shown in references (McCoy, 1932; Cohen, 1966; Summerfield and
Zweifel, 1969; Cohen, 1966, 1976, 1987), these quasi-distribution
functions can be obtained from the \wqdf\ by simple
transformations.

The equality derived in the Appendix A is applicable whenever the
operators obey the relation
\eqn{
\hat P_{can}=\hat P_W +A(\hat X),}
{rel}
where $\hat P_W$ is the desired variable in the Wigner
representation. That is, as long as one chooses a variable that
obey (\ref{rel}), the transformation (\ref{Fourier3}) should be
used.

In the language of classical mechanics it can be stated that the
relation (\ref{rel}) holds if the coupling to the electromagnetic
field is via the term $A(X)\dot X$ in the Lagrangian $L$. Then the
canonical momentum is of the form
$$P_{can}=$$
\eqn{\frac{\pd L(X,\dot X,A)}{\pd \dot X}= \left.\frac{\pd
L(X,\dot X,A)}{\pd \dot X}\right|_{A=0}+ A(X)\stackrel{\rm def}{=}
P_W +A(X).} {rrel}
Whenever an effective Hamiltonian is used the relation (\ref{rel})
is assumed to hold (see, e.g., Nenciu, 1991, and references
therein). This however, is not necessarily the case. The particle
may have a different dispersion relation due to its coupling to a
set of degrees of freedom (e.g., polaron). Then if the Hamiltonian
for such a composed quasi-particle is written, the minimal
coupling (\ref{rel}) may be not necessarily applicable.

\subsection{Kinetic and canonical momenta in the Wigner representation}

The phase factor in the \wqdf , equation (\ref{wqdf2}), can be
written as a Taylor series and integrated over $s$ so that
equation (\ref{wqdf2}) takes the form
\begin{eqnarray}
\rho_W(P,X) & = & \int \di y \exp\left({-i \over \h}  y^\mu
\left[P_\mu - {e\over c} j_0(y^\nu \partial_\nu^X/2) A_\mu (X)
\right]\right)\times \nonumber
\\ & & \left\langle \Psi^\deg(X+y/2)\Psi(X-y/2) \right\rangle
\nonumber
\\
&= & \exp\left( {e\over c }j_0 \left( {\hbar\over 2}
\partial^{P\nu}\partial^X_\nu
\right) A_\mu(X) \partial^{P\mu} \right)\times \nonumber
\\
& & \int \di y \exp\left({-i \over \h}  y^\mu P_\nu\right)
\left\langle \Psi^\deg(X+y/2)\Psi(X-y/2) \right\rangle \nonumber
\\
& = & \exp\left( {e\over c}\slashbA_\mu (X)
\partial^{P\mu}\right) \rho_{can}(P,X) \label{wqdf6}
\end{eqnarray}
\marginpar{wqdf6}
\noindent where $\slashbA_\mu (X) =  j_0(\Delta/2) A_\mu (X)$,
$\Delta = \hbar\partial^{P\nu} \partial^X_\nu$, $j_0(x) = \sin x/x
\approx 1 - x^2/3 + \cdots$, and the derivatives with respect to
$X$ in the exponent act only on the vector potential.

This is a general expression valid for an arbitrary
electromagnetic field. We define $\tilde\rho_W(P,X)$ and $\tilde
P$ to be the \wqdf\ and its  momentum as derived from equation
(\ref{wqdf6}) with the modified phase factor $\exp\left({e \over
c} \left[ \slashbA_\mu (X) - A_\mu (X) \right]
\partial^{P\mu} \right)$. The equality
\eqn{
\rho_W(P,X)=\tilde\rho_W \left( \tilde P-{e\over c}A(X),X \right)
}{wqdf12}
then holds and one may say that the two momenta are related by the
equation
\eqn{
P_{kin}(X)=\tilde P(X) -{e\over c}A(X).
}{wqdf0}
If all the derivatives of the electromagnetic field of order
higher than one vanish, meaning only static and homogeneous fields
are applied, the relation (\ref{wqdf0}) is simplified and $\tilde
P$ becomes the usual canonical momentum. The kinetic momentum and
the canonical momentum as defined by the \wqdf s in this section
are related by the equality,
\eqn{ \left. P_{can}\right|_{X=x''} =
P_{kin}(X)=P_{can}(X;x'')-(e/c)A(X;x'') } {wqdf5}
where $A(X;x'')$ is in the axial gauge, (\ref{int22}).

The differential version of equation (\ref{wqdf5}) is,
\eqn{
\partial^X_\mu P_{kin}(X) =\partial^X_\mu P_{can}(X;x'')+
\partial^{x''}_\mu P_{can}(X;x'')
.}{difff}
Equation (\ref{difff}) shows us that the variation of the
canonical momentum is composed of two parts: variation of the
kinetic momentum due to kinetic effects and variation of the
kinetic momentum due to the change of the gauge as the reference
point $x''$ moves. In particular we may use equation (\ref{int22})
and write (for systems in static and homogeneous field)
\eqn{
\frac{\di}{\di t''} P_{can}(X;x'') ={e\over 2} \bE+{e\over 2c}
\frac{\di \bx''}{\di t''}\times\bH= {1\over 2}\bFl_{Lor}(x''),
}{dif}
where $\bFl_{Lor}(x'')$ is the Lorentz force acting at the point
$x''$.

Let us imagine, a quantum system confined to a small part of the
three dimensional space such that it can be described as being at
rest (equivalently we may think of a heavy mass particle). Another
classical particle with a well-defined trajectory is chosen to
define the location of the point $x''$. The gauge part of the
canonical momentum of the quantum system (as measured in the axial
gauge by $x''$ at the classical particle) will change in time
according to the equation of motion of the classical particle
(\ref{difff}). While all that time, the kinetic part of the
kinetic momentum of the quantum system will not change since it is
at rest. We shall return to this point twice, when discussing the
heavy mass polaron and when discussing the `acceleration theorem'.

Now, suppose that the classical particle is another quantum system
that interacts with the first system and exchanges momenta with
it. Momentum conservation applies only to the canonical momentum.
If the conservation laws are to be expressed in terms of kinetic
momenta, the conservation laws should take into account the
kinetic as well as the gauge part of the kinetic momenta. More on
this point is discussed in Subsection \ref{KQ}.
\subsection{The classical limit of the \wqdf.}
The classical limit of the \wqdf\ is discussed here. This limit
has been already addressed by other authors (Heller, 1976, 1977;
Berry, 1977) and here we give only the modifications due to the
gauge invariance and appearance of an energy variable. A different
approach was taken by Bund (1995).

Following Van Vleck (1928) we express the quasi-classical limit of
a wave function for a state $\alpha$ as
\eqn{
\psi_\alpha(X) = N^{1/2} {\cal D}^{1/2} e^{S/i\hbar}.
}{vv}
$N$ is the normalization constant, $S(X,\alpha) = S({\bf
R},t,\alpha)$ is the classical generator of the canonical
transformation from the ${\bf P}$, ${\bf R}$ system to a set of
new variables $\alpha$, $\beta$ and ${\cal D} = \det|\partial^2
S/\partial {\bf R}\partial\alpha |$ is the functional determinant.
The \wqdf\ reads

$$\rho_W(P,X|\alpha) =$$
$$\int \di^4 y
\sqrt{\det|\partial^2 S(X+y/2,\alpha)/\partial {\bf R}\partial\alpha|
\det|\partial^2 S(X-y/2,\alpha)/\partial {\bf R}\partial\alpha|}\times
$$
\eqn{
\exp\left[{-i\over\hbar}\left[
(P-A(X))^\mu y_\mu + S(X-y/2,\alpha) -S(X+y/2,\alpha)\right]
\right].
}{ghj}
We neglected terms, non-singular in $\hbar$, in the phase factor,
since they vanish as $\hbar\rightarrow 0$.

We conclude that the relation (\ref{wqdf5}) is expected to hold in
the quasi-classical limit as well as for quantum systems in static
and homogeneous fields. In the classical limit the frequency and
the wave length of the electron wave function are much smaller
than any other time and space variations in the system, that is,
the electromagnetic field is taken as static and homogeneous. This
relation between the canonical and kinetic momenta are always
valid in classical mechanics as well as for quantum mechanical
operators. However, in the Wigner representation of the quantum
mechanics this relation generally does not necessarily hold.

\section{Derivation of kinetic equations.}
\label{kineticequations}
The diagrammatic technique proposed by Keldysh (1965) allows one
to write the Dyson equation for a nonequilibrium quantum system in
an arbitrary not quantized electromagnetic field. A semiclassical
Boltzmann equation as well as its quantum versions may be obtained
from this Dyson equation by making certain transformations and
approximations. It is our aim to present below quantum equations
avoiding approximations as much as possible. In this sense, both
the equations to be obtained and the Dyson equations are equally
valid.

Following the five step procedure presented in our paper (Levanda
and Fleurov, 1994) we arrive at equations which are quantum
analogs of the Hamilton-Jacobi and the Boltzmann kinetic equations
$$
{2\over \hbar}\left[\slashvarepsilon - {1\over 2m}{\slashbP}^2 +
{\hbar^2\over 8m}(\slashnabla^{X})^2\right]
\hat \bG(P,X)= 2 \sigma_x
$$
$$
+
\exp\left[{i\hbar\over 2} \left(
\frac{\partial}{\partial \varepsilon}\slashpartial^T  -
\slashpartial^T \frac{\partial}{\partial\varepsilon} +
\slashnabla^x \frac{\partial}{\partial\bP} -
\frac{\partial}{\partial\bP}\slashnabla^x
\right)\right]
$$
\eqn{ \left( \sigma_x \hat\bSigma(P,X) \hat\bG(P,X)+ \hat\bG(P,X)
\hat\bSigma(P,X) \sigma_x \right) }{qbehj}
and
$$
i\left[\slashpartial^T + {1\over 2 m}
\left({\slashbP}
\cdot\slashnabla^X +\slashnabla^X \cdot{\slashbP}\right)
\right]
\hat \bG(P,X)=
$$
$$
\exp\left[{i\hbar\over 2} \left(
\frac{\partial}{\partial \varepsilon}\slashpartial^T  -
\slashpartial^T \frac{\partial}{\partial\varepsilon} +
\slashnabla^x \frac{\partial}{\partial\bP} -
\frac{\partial}{\partial\bP}\slashnabla^x
\right)\right]
$$
\eqn{
\left( \sigma_x
\hat\bSigma(P,X) \hat\bG(P,X)-
\hat\bG(P,X) \hat\bSigma(P,X)
\sigma_x \right).
}{qbebl}
where the following definitions are used
\begin{eqnarray*}
\slashvarepsilon &=& \varepsilon - {\hbar\over 2}ej_1\left(
{\Delta\over 2}\right)\bE(X)\cdot\bnabla^P; \\
{\slashbP} &=& \bP + {\hbar\over 2}ej_1\left(
{\Delta\over 2}\right)
\left({1\over c}\bB(X)\times\bnabla^P + \bE(X)\partial^\varepsilon
\right); \\
\slashpartial^T &=& {\partial\over\partial T}
+ej_0\left({\Delta\over 2}\right)
\bE(X)\cdot\bnabla^P; \\
\slashnabla^X &=& \bnabla^X +ej_0\left({\Delta\over 2}\right)
\left({1\over c}\bB(X)\times\bnabla^P + \bE(X)\partial^\varepsilon
\right);
\end{eqnarray*}
\eqn{
\hat\bG= \left(
\begin{array}{cc}
 0 & G^a \\
 G^r & G^K
\end{array}
\right) \; \; \;
\hat\bSigma= \left(
\begin{array}{cc}
\Sigma^K & \Sigma^r \\
\Sigma^a & 0
\end{array}
\right);
}
{Green2}
$$j_0 (x)=\sin(x)/x\ \ \ \mbox{and}\ \ \ j_1(x) = \sin(x)/x^2 -
\cos(x)/x.$$ Here the derivative with respect to $X$ in the
operator $\Delta ={\h}\partial^P\partial^X $ acts only on the
electromagnetic fields ${\bf E}$ and ${\bf B}.$ The other
derivatives, in the left hand sides, with respect to $X$ and $P$
act on all the functions. In the right hand side, the first and
the second derivatives of in the exponent act on the first or on
the second terms of the pair in the sum, respectively.

This is a generalization of the so-called gradient expansion first
used by Kadanoff and Baym (1962) for system without
electromagnetic fields, by Tugushev and Fleurov (1983) for system
in homogeneous and static electric field and by Mahan (1987) for
homogeneous and static electric and magnetic fields. The form
(\ref{qbehj}), (\ref{qbebl}) is more general and valid for
arbitrary electromagnetic fields. An alternative representation of
the gradient expansion was given by these authors (Levanda and
Fleurov, 1994) in terms of the phase loop function (see also
Appendix B).

These equations are equivalent to the Dyson equations for the
Keldysh functions from which we have started. In fact, they are
just alternative representations of the Dyson equations. Therefore
they may provide an accurate and complete physical description for
kinetic processes of a many particle system in an arbitrary
electromagnetic field. We wrote two matrix equations which would
correspond to six equations for complex functions. However, the
number of really independent functions and, hence independent
equations is only two (see, e.g., discussion in Fleurov and
Kozlov, 1978). One can, for example, take an off-diagonal term in
the matrix equation (\ref{qbehj}) in the representation
(\ref{Green2}) which would be an equation for, say, retarded
Green's function $G^r$ determining the spectrum of the system. The
second equation can be produced from one of the diagonal terms of
the matrix equation (\ref{qbebl}) (in the same representation) and
this is the equation for the function $G^K$ which in the
semi-classical approximation reduces to the conventional Boltzmann
equation.

Both equations can be expanded in powers of the derivatives of the
electromagnetic fields ($\Delta$ expansion), without effecting
their gauge invariance. This, however, does not correspond to the
expansion in powers of $\hbar$  of equation (\ref{qbehj}),
(\ref{qbebl}).

The Liouville theorem in classical mechanics can be derived as a
limit of equation (\ref{qbebl}). To see it we neglect the
interaction term (the right hand side) and  integrate with respect
to the variable $\varepsilon$ (the classical phase-space does not
contain the energy variable). Equation (\ref{qbebl}) becomes
\eqn{
\left[
{\partial\over\partial T} +
ej_0\left({\Delta\over 2}\right)
\left[{\bf E}(X) +
{1\over m}
{\bf P}\times {1\over c}{\bf B}(X)
\right]\cdot \bnabla^P
+{\bf P}\cdot\bnabla^X
\right]
\hat \bG(\bP,X)= 0.
}{clss}
If all the derivatives of the fields of order higher than one,
vanish, equation (\ref{clss}) becomes classical Liouville
equation. This should be compared with:
\begin{enumerate}
\item The conditions for the coincidence of the Ehrenfest
equations of motion for a wave packet in an electric field (see,
e.g., Messiah, 1962, p.220) with the classical equation of motion
(if the electric field has at most a linear coordinate
dependence).
\item The conditions for a quantum system to be deterministic.
Moyal (1949) showed that if the Hamiltonian of a quantum system is
a second-degree polynomial in $X$ and $P$ (a uniformly accelerated
particle or a harmonic oscillator), then the quantum system is
deterministic in the sense that the time evolution of the \wqdf\
is given by the classical Liouville theorem. That is, the quantum
nature of the system enters only via the initial conditions
(Bartlett and Moyal, 1949; Lee and Scully, 1983).
\end{enumerate}

\subsection{Kinetic meaning of the slashed terms
\label{slashsec}}

We shall now demonstrate the relation of the slashed energy,
momentum and derivatives to the operators of the kinetic energy
and kinetic momentum. This will clarify the physical significance
of these quantities and their difference from the non-slashed
quantities. We also give two examples of the \wqdf\ that
demonstrate the difference between the two derivatives.

The slashed time derivative is defined as
\begin{eqnarray}
\lefteqn{
i\h\slashpartial^T G^<(P,X)=\nonumber} \\
i\h\slashpartial^T
&&
\dint \di^4 y \exp\left( -{i\over\h} P y +
{i e\over c\h} \dint^{1/2}_{-1/2} y^\mu A_\mu
(X+sy)\di s
\right) G^<(y,X)
\nonumber \\
&=&
\dint \di^4 y \exp\left( -{i\over\h} P y +
{i e\over c\h} \dint^{1/2}_{-1/2} y^\mu A_\mu
(X+sy)\di s
\right)\cdot
\nonumber \\
&&
 \left[i\h{\partial\over\partial T} + {e\over c}A_0(X-y/2)-
{e\over c}A_0(X+y/2) \right] G^<(y,X) \nonumber \\ &=& \dint \di^4
y \exp\left( -{i\over\h} P y + {i e\over c\h} \dint^{1/2}_{-1/2}
y^\mu A_\mu (X+sy)\di s \right)\cdot \nonumber \\ &&
\left[\left(i\h{\partial\over\partial t} + {e\over c}
A_0(x)\right)- \left(-i\h{\partial\over\partial t^\prime} +
{e\over c}A_0(x^\prime)\right) \right] i\langle
\psi^+(x^\prime)\psi(x) \rangle \label{Fourier9}
\end{eqnarray}
This definition has been written for the $G^<(x,x^\prime) =
i\rho(x,x^\prime)$ component. The expressions for the other
components are similar. The definition of the slashed spatial
derivative is
$$
-i\h\slashnabla^X G^<(P,X)=
\dint \di^4 y \exp\left( -{i\over\h} P y +
{i e\over c\h} \dint^{1/2}_{-1/2} y^\mu A_\mu
(X+sy)\di s
\right)\cdot
$$
\eqn{ \left[\left(-i\h\bnabla^x  - {e\over c}\bA(x)\right)-
\left(i\h\bnabla ^{x^\prime}- {e\over
c}\bA(x^\prime)\right)\right] i\langle \psi^+(x^\prime)\psi(x)
\rangle }{Fourier10}

In a similar way one can introduce definitions for the slashed
energy and momentum
$$
2
\slashvarepsilon G^<(P,X) =
\dint   \di^4 y \exp\left( -{i\over\h} P y +
{i e\over c\h} \dint^{1/2}_{-1/2} y^\mu A_\mu
(X+sy)\di s
\right)\cdot
$$
\eqn{\left[\left(i\h{\partial\over\partial t} + {e\over
c}A_0(x)\right)+ \left(-i\h{\partial\over\partial t^\prime} +
{e\over c}A_0(x^\prime)\right)\right]
i\langle\psi^+(x^\prime)\psi(x) \rangle }{Fourier19}
$$ 2\slashbP G^<(P,X)= \dint   \di^4 y \exp\left( -{i\over\h} P y
+ {i e\over c\h} \dint^{1/2}_{-1/2} y^\mu A_\mu (X+sy)\di s
\right)\cdot $$
\eqn{
\left[\left(-i\h\bnabla^x  - {e\over c}\bA(x)\right)
+ \left(i\h\bnabla ^{x^\prime}-
{e\over c}\bA(x^\prime)\right)\right]
i\langle\psi^+(x^\prime) \psi(x) \rangle
}{Fourier29}
The operators in the square brackets in (\ref{Fourier19}) and in
(\ref{Fourier29}) are the kinetic energy and kinetic momentum
operators. The kinetic momentum operator that acts on the wave
function of the system is gauge invariant and produces the values
of the electron's mass times a quantity which has a meaning of an
average velocity.

We shall now show how the uniformity of the particle and the
current density make $\slashbnabla^X \hat\bG(P,X)$, but not
$\bnabla^X \hat\bG(P,X)$, vanish. The system is not assumed to be
homogeneous, e.g., it may be placed in a non-uniform magnetic
induction or in a time dependent electric field. That is, we
assume only that the density and the current do not depend on the
coordinate:
$$
\psi^*(x) \psi(x)= \rho(x) = \rho
$$
and
$$
{e\over 2m}\left.\left[\left(-i\h\bnabla^x  - {e\over c}\bA(x)\right)
+ \left(i\h\bnabla ^{x^\prime}-
{e\over c}\bA(x^\prime)\right)\right]
i\psi^*(x^\prime) \psi(x) \right|_{x^\prime\rightarrow x}
= \bj(x)=\bj
$$
and show that $\slashbnabla^X \hat\bG(P,X)=0$.

A single particle `wave function' of the system is written in the
form $\psi(x) = R(x)\exp[iS(X)/\hbar]$ and with the two
aforementioned conditions for homogeneity, one finds, $R(x)=
\sqrt{\rho}$ and $\bnabla^x S(x) = m\bj/e + \bA(x)$. Putting this
form of the wave function into  equation (\ref{Fourier10}) the
derivative $\slashbnabla^X \hat\bG(P,X)$ is readily seen to
vanish. The derivative $\bnabla^X \hat\bG(P,X)$ vanishes as well,
only if the electromagnetic fields vanish. Substituting the same
wave function in equation (\ref{Fourier29}) we find that $\slashbP
G^<(P,X)= m\bj/e G^<(P,X)$ and $\bP G^<(P,X)\neq m\bj/e G^<(P,X)$
if the electromagnetic fields have non-vanishing gradients.

The \wqdf\ for a charged particle in a static and homogeneous
electric field in a system without interactions obey the equations
\eqn{
\left[ \varepsilon - {\bP^2 \over 2m} +
{\h^2 \over 8 m}\left(
\bnabla^X +
e\bE {\partial^\varepsilon} \right)^2 \right]
\rho_W (P,X) = 0
}{grd}
\eqn{
\left[{\partial\over\partial T} +
e\bE\bnabla^P+
{\bP\over m} \left(\bnabla^X+
e\bE{\partial^\varepsilon}\right)\right]
\rho_W(P,X)=0
}{gre}
If the condition is assumed that the derivatives of \wqdf with
respect to the coordinates, $\bnabla^\bX\rho(P,X)=0,$ vanish then
the \wqdf\ reads
$$\rho_W(P,X)=$$
\eqn{{m\over (m\hbar e E)^{2/3}} {\rm Ai}\left( {2m\over (m\hbar e
E)^{2/3}} \left( \varepsilon - {\bP^2\over 2m} \right) \right)
\rightlimit{\hbar\bE\rightarrow 0} \delta \left( \varepsilon -
{\bP^2\over 2m} \right) }{airyw}
which is a time and coordinate independent solution, ${\rm Ai}$ is
the Airy function. The solution (\ref{airyw}) implies that the
application of an electric field accelerates the particle and,
hence, causes a broadening of the energy uncertainty. The
broadening is of the order of $(\hbar\bE)^{2/3}$ and is only due
to the acceleration, no interactions are assumed. The \wqdf\ is a
spectral function that yields an electron dispersion relation. It
does not produce the dynamics of the system. The dependence on the
energy and the momentum can be rewritten in the form
$$ \varepsilon - {\bP^2\over 2m} = \varepsilon - e\bE \left(
{\bP\over m}T - {e\bE\over 2m}T^2\right) - {\left(\bP - e\bE
T\right)^2\over 2m}= \varepsilon - e\bE \bR(T) - {\bP^2(T)\over
2m}, $$
where $\bR(T)$ and $\bP(T)$ are the coordinate and the momentum of
a classical particle that moves in the same electric field. This
representation reveals the hidden acceleration of the particle. If
a steady current flows in the system the slashed derivatives
vanish and the dispersion relation found from equation (\ref{grd})
is $\delta \left(\varepsilon - {\bP^2\over 2m}\right)$, hence, no
uncertainty broadening takes place. We conclude that the \wqdf\ of
a system without interactions that does not depend on the
coordinates correspond to a particle accelerated in the electric
field and not to a steady current system.

Now turning to a system with interactions we may use at this stage
the simplest mean free time approximation in the quantum kinetic
equations. This approximation has been used for more than a
century in classical and quasi-classical systems (see, e.g.,
Ziman, 1960). To introduce the mean free time we first consider a
solution without any field $\left. \rho(P,X) \right|_\bE = 0$.
Then the Liouville equation (\ref{clss}) allows as to make the
`canonical transformation'
\begin{eqnarray*}
\varepsilon &\rightarrow &\varepsilon - e\bE \cdot (\bR -\bR_0), \\
\bP & \rightarrow & \bP-e\bE  T, \\
\bR & \rightarrow & \bR_0 - {\bP\over m} T +{e\over 2m} \bE T^2.
\end{eqnarray*}
in order to arrive to a solution in an electric field. The
transformed \wqdf\ after the canonical transformation has the
following properties: (i) If the system, has been homogeneous
prior to the application of the electric field, it remains such
afterwards; (ii) The slashed derivative vanishes.

The next step is to assume that after the interaction converts the
variable $T$ into a parameter having the meaning of the mean free
time that needs to be determined from the quantum kinetic
Boltzmann equation. The mean free time may be a function of the
momentum and the energy, $\tau(\varepsilon,\bP)$. The derivative
with respect to $T$ in the kinetic equations vanishes ($T$ is now
a parameter) as well as the derivatives with respect to $X$ and
$\varepsilon$ (due to the canonical transformation of
$\varepsilon$). In the left hand side of the Boltzmann kinetic
equation (\ref{gre}) we are left with only one term,
$e\bE\cdot\bnabla^\bP\rho_W(\varepsilon,\bP,\bR)$, just like in a
quasi-classical Boltzmann equation (see, e.g. Ziman 1960). This
term should be equal to the collision integral in the right hand
side of the equation, the electric force is equal to the collision
force. In the left hand side of the Hamilton-Jacobi kinetic
equation (\ref{grd}) we are left with $(\varepsilon -
P^2/2m)\rho_W(\varepsilon,\bP,\bR)$ that is equal to the
interaction which causes a broadening in the dispersion relation
and vanishes for classical systems.

Let us look at the dependence of the \wqdf\ on the mean free time.
The \wqdf\ in equilibrium is usually written in the form $\left.
\rho_W(P,X)\right|_{\bE=0} = A(P,X) n_F(\varepsilon)$ where
$n_F(\varepsilon)$ is the Fermi function and $A(P,X)$ is the
spectral function. In the linear approximation the collision
integral does not vanish only if nonequilibrium corrections to the
Fermi function are considered, the spectral function may be taken
at equilibrium (Fleurov and Kozlov, 1978). Hence, the linear
correction to the \wqdf\ is taken in the form
\begin{eqnarray}
\rho_W(P,X) &= &A(P,X)\left[ n_F(\varepsilon -
e\bE \cdot (\bR -\bR_0))\right]\nonumber \\
&\approx &
A(P,X)\left[ n_F(\varepsilon) -
\left(\partial n_F\over\partial\varepsilon\right)
e\bE \cdot (\bR -\bR_0)\right]\nonumber \\
&=&
\left.\rho_W(P,X)\right|_{\bE=0} +
A(P,X)\left(\partial n_F\over\partial\varepsilon\right)
{e\bE \cdot \bP\over m}\tau(\varepsilon,\bP,\bR).
\end{eqnarray}
This form of the \wqdf\ may be inserted into the quantum kinetic
equation in order to calculate the mean free time
$\tau(\varepsilon,\bP,\bR)$.

\section{Kinetic momentum continuity equation \label{AB}}
Quantum kinetic equations provide us a possibility to calculate
Green's functions. Green's functions, however, are not directly
measurable, and are only used at certain steps of calculations.
Green's functions enable one to calculate quasi-distribution
functions, and macroscopic measurable quantities are calculated as
their averages over the phase space. Although deduction of various
equations for the macroscopic measurable quantities from the
quantum kinetic equations is possible, we shall consider here only
the {\em kinetic momentum continuity equation}. The {\em particle
continuity equation} and the quantum mechanical equivalent of the
classical {\em Hamilton-Jacobi equation} were discussed in Levanda
and Fleurov (1994).

\subsection{Space conditional moments}

Macroscopic measurable quantities can be written in terms of gauge
invariant space conditional moments, $\dbarPn_\mu (X)$ which can
be defined as the averages of various powers $(P_\mu)^n$ of the
momentum at a given point, $X$,
\begin{eqnarray}
\lefteqn{
\dbarPn_{\mu}(X) = \frac{1}{\rho (X)}
{1\over (2\pi\hbar)^4}
\dint \di^4P (P_\mu)^n \rho_{W}(P,X)
\nonumber} \\
&=&
\frac{1}{\rho (X)}
\left(-i\hbar \frac{\partial}{\partial y}\right)^n
\left[
\exp\left({i e\over c\h} \dint^{1/2}_{-1/2} y^\mu A_\mu
(X+sy)\di s\right)
\rho(X,y)
\right]_{y\rightarrow 0}.
\label{conm}
\end{eqnarray}
Here
$$\rho (X) =  {1\over (2\pi\hbar)^4}\dint \di^4 P \rho_W(P,X)$$
is the particle density. The canonical (non gauge-invariant) space
conditional moments are defined as
$$\dbarPn_{can,\mu}(X) =$$
\eqn{
 \frac{1}{\rho (X)}{1\over (2\pi\hbar)^4}
\dint \di^4P (P_\mu)^n \rho_{W}(P_{can},X)=
\frac{1}{\rho (X)}
\left(-i\hbar \frac{\partial}{\partial y}\right)^n
\rho(X,y)
}{btf}

The relation between the canonical space conditional moments and
gauge-invariant space conditional moments is trivial if equation
(\ref{wqdf5}) holds. In the general case one can also see that the
relation
\eqn{
\dbarP(X;x'')=\dbarP_{can}(X) -(e/c)A(X;x'')}{pp}
holds regardless of the nature of the electromagnetic field (as
well as equation (\ref{wqdf0})).

This can be readily understood, since $\dbarP_{can}(X)$ is merely
the expectation value of the current operator times the electron
mass divided by the density,
$$\dbarP_{can}(X) = \frac{m \dbarJ_{can}(X)}{e\rho (X)} = $$ \eqn{
\left. {-i\h\over 2}
\frac{\left(\partial^{X'}-\partial^X\right)\left\langle
\Psi^\deg(X')\Psi(X)\right\rangle}
{\left\langle\Psi^\deg(X')\Psi(X)\right\rangle}
\right|_{X'\rightarrow X}. }{curr1}
In a similar manner we have an expression for the gauge-invariant
terms
$$ \dbarP(X)= \frac{m \dbarJ (X)}{e\rho (X)}=$$ \eqn{ \left. {1
\over 2} \frac{\left(-i\h
\partial^{X^\prime}-{e\over c}A(X^\prime)+ i\h \partial^X +
{e\over c}A(X) \right)\left\langle
\Psi^\deg(X')\Psi(X)\right\rangle}
{\left\langle\Psi^\deg(X')\Psi(X)\right\rangle}
\right|_{X'\rightarrow X}. }{curr2}
Relations between the canonical and gauge-invariant conditional
moments are best seen with the help of the cummulant expansion
generated by the function
$$ \log\left( \exp\left({i e\over c\h} \dint^{1/2}_{-1/2} y^\mu
A_\mu (X+sy)\di s\right) \rho(X,y)/\rho (X)\right) $$
$$ = {i e\over c\h} \dint^{1/2}_{-1/2} y^\mu A_\mu (X+sy)\di s +
\log\left(\rho(X,y)/\rho (X)\right). $$
Then writing the density function in the form
$$ \rho(X,y)= \rho(X-y/2)\rho(X+y/2)\exp {-i\over\hbar}\left(
S(X-y/2)-S(X+y/2) \right)$$
one obtains cummulants.
$$\kappa_{2n+1}=$$
\begin{eqnarray*}
& \left({-i\hbar\over 2}\right)^{2n} \left({\partial\over\partial
X}\right)^{2n+1} S(X) + \left(-i\hbar \frac{\partial}{\partial
y}\right)^{2n+1} \left[ {i e\over c\h} \dint^{1/2}_{-1/2} y^\mu
A_\mu (X+sy)\di s \right]_{y\rightarrow 0} &=\\ &
\left({-i\hbar\over 2}\right)^{2n} \left({\partial\over\partial
X}\right)^{2n+1} S(X) + \left({-i\hbar\over 2}
\frac{\partial}{\partial X}\right)^{2n} {e\over c} A(X)&
\end{eqnarray*}
and
\begin{eqnarray*}
\kappa_{2n}&=& \left({-i\hbar\over 2}\right)^{2n}
\left({\partial\over\partial X}\right)^{2n} \log\rho(X)
+
\left(-i\hbar \frac{\partial}{\partial y}\right)^{2n}
\left[
{i e\over c\h} \dint^{1/2}_{-1/2} y^\mu A_\mu
(X+sy)\di s
\right]_{y\rightarrow 0}
\\
&=& \left({-i\hbar\over 2}\right)^{2n}
\left({\partial\over\partial X}\right)^{2n} \log\rho(X).
\end{eqnarray*}
$n=0,1,2 \ldots$ The odd cummulants contain canonical cummulants
plus cummulants of the phase factor. The asymmetry of the \wqdf\
is, hence, represented by these odd cummulants, therefore the
asymmetry of the gauge-invariant and canonical momenta are
different. The even gauge-invariant and canonical cummulants are
the same and, in particular, the conditional mean square deviation
of the momenta, $\kappa_2= \overline{\overline{{\bf P}^2}}-
{\overline{\overline{{\bf P}}}}^2,$ is the same in both cases and
the uncertainty relation holds for both the canonical and the
gauge-invariant momenta.

\subsection{Derivation of kinetic momentum continuity equations}
Here we discuss an equation which can be called a kinetic momentum
continuity equation. It is obtained from the trace of equation
(\ref{qbebl}), after multiplying it by ${\bf P}$ and integrating
over the energy and momentum,
$$
\frac{\partial}{\partial T}\left(\rho (X)\dbarbP (X)\right)
+ \bnabla^{X\lambda}\left(\frac{\dbarPPl (X)}{m}\rho (X)\right)
=
$$
$$-e\bE (X)\rho (X)-(e/mc)\left(\rho (X)\dbarbP
(X)\right)\times\bB(X)+$$
\eqn{ {1\over (2\pi\hbar)^3}\dint \di \bP \;\bP I_{coll} (\bP,X).
} {mce}
The kinetic momentum continuity equation (\ref{mce}) contains a
collision integral, multiplied by ${\bf P}$ and integrated over
the momentum space. It describes the transfer of electron kinetic
momentum from and to a four dimensional volume element, at $X$,
caused by the collision processes.

Equation (\ref{mce}) can be rewritten in a more transparent form
$$m\frac{\di\dbarv (X)}{\di T} =$$
\eqn{ -e\bE (X)-{e\over c}\dbarv (X) \times\bB(X)
-\frac{\h^2}{2m}\bnabla \frac{\bnabla^2\sqrt{\rho(X)}}
{\sqrt{\rho(X)}} +{\bf f}_{coll}(X) }{mce2}
where
$$ {\bf f}_{coll}(X)= {1\over \rho (X)} {1\over (2\pi\hbar)^3}
\dint \di \bP \;(\bP-{m\dbarv (X)}) I_{coll} (\bP,X), $$
$$\dbarv (X)= \dbarbP (X)/m=
\dbarJ_{kin}(X) /e\rho (X)$$
and the equality
$$-\rho(X)\frac{\h^2}{2m}\bnabla \frac{\bnabla^2\sqrt{\rho(X)}}
{\sqrt{\rho(X)}} = \bnabla^\lambda \sigma_{\lambda\nu} =$$
\begin{equation}
\bnabla^\lambda\left[\rho(X)\left(\dbarPPl (X)
-\dbarP_\lambda(X)\dbarbP(X) \right)\right] \label{qupot}
\end{equation}
is used. The left hand side of equation (\ref{qupot}) is the
gradient of the so-called `quantum potential', whereas the term
$\sigma_{\lambda\nu}= (\h^2/4m)\rho(X)\bnabla^\lambda \bnabla^\nu
\ln\left(\rho(X)\right)$ is the stress tensor used in the
hydrodynamic formulation of the Schr\"odinger equation (Madelung,
1927; Holland, 1993). The electron density continuity equation
(equation (47) in Levanda and Fleurov, 1994) shows that the
quantity $e\rho(X)\dbarbP (X)/m$ is the kinetic particle current
density $\dbarJ_{kin}(X)$ at the point $X$, hence, $\dbarv (X)$ is
a quantity having a meaning of an average velocity of the
particles. This averaging includes both thermodynamical and
quantum averagings.

The collision term ${\bf f}_{coll}(X)$ appears due to possible
processes of electron scattering. Bohm (1952) (his equation (31))
argued that hydrodynamic equation in the quantum theory should
contain terms depending on the difference ${\bf p}-m\dbarv$ but
did not present a physical mechanism that may lead to such a term.
It is clear that the collision term  ${\bf f}_{coll}(X)$ is a
function just of this type. Its specific form is determined by the
collision integral and can be now calculated explicitly for any
particular system.

This continuity equation has been obtained from the gauge
invariant quantum kinetic equations which depend explicitly only
on electromagnetic fields and do not depend on their potentials.
These equations are actually Dyson equations for the Keldysh
Green's functions in a differential form. In the absence of
collision processes, they can be reduced to a Schr\"odinger
equation and to a quantum generalization of the Liouville theorem
(Moyal, 1949). All the physical processes described by the
Schr\"odinger equation can be described by the quantum kinetic
equations without any need of using potentials or
vector-potentials of the field. The continuity equation for the
velocity $\dbarv (X)$ can be looked at as a {\em quantum
generalization of the second law of Newton}. Equation (\ref{mce2})
converts into a classical Newton equation of motion with a Lorentz
force, if the quantity $\dbarv (X)$ is replaced by the classical
kinetic velocity and $\h\rightarrow 0$.

We mention, here, a few general things about the collision terms
that emerge from their explicit expressions derived from equation
(38) of Levanda and Fleurov (1994). The collision force is written
in the form
$$  \rho(X) f_{coll}(X)=$$
\eqn{\left. -i\hbar\bnabla^y \tilde D(X,y)\right|_{y\rightarrow 0}
\left. - \left[m\dbarv(X)+ {e\over c}A(X)\right]\tilde
D(X,y)\right|_{y=0}}{colforc}
where
$$
\left.\tilde D(X,y)\right|_{y_0=0}=
\int \di^4 x_1 \theta_H \left(X_0 - (x_1)_0\right)
\left.\tilde D(X,y|x_1)\right|_{y_0=0}
$$
and
\begin{eqnarray*}
\left.\tilde D(X,y|x_1)\right|_{y_0=0}&=&
\left[
\Sigma^>(X_0,\bX-\by,(x_1)_0,\bx_1)
G^<((x_1)_0,\bx_1,X_0,\bX+\by)\right.
\\
&&
-
\Sigma^<(X_0,\bX-\by,(x_1)_0,\bx_1)
G^>((x_1)_0,\bx_1,X_0,\bX+\by)
\\
&&
-
G^>(X_0,\bX-\by,(x_1)_0,\bx_1)
\Sigma^<((x_1)_0,\bx_1,X_0,\bX+\by)
\\
&&
+
\left.
G^<(X_0,\bX-\by,(x_1)_0,\bx_1)
\Sigma^>((x_1)_0,\bx_1,X_0,\bX+\by)
\right].
\end{eqnarray*}
If the Green's functions and the mass operators obey the
equalities $\tilde G^<(X,x_1)= - \tilde G^>(x_1,X)$ and
$\tilde\Sigma^<(X,x_1)= - \tilde\Sigma^>(x_1,X)$, then $\tilde
D(X,y|x_1) = \tilde D(X,-y|x_1)$ and all the processes in the
system are detailed balanced as discussed in Section 4 of Levanda
and Fleurov (1994). In particular creation and annihilation of
particles, $\left.\tilde D(X,y)\right|_{y=0}=0,$ and the collision
force, $\bnabla^y\left.\tilde D(X,y)\right|_{y\rightarrow 0}=0$,
vanish.

If transport processes occur in the system, the detailed balance
no longer exists and the conservation of the number of particles
or a vanishing collision force are not necessarily true. The
function $\left.\tilde D(X,y|x_1)\right|_{y=0}=0$ can be regarded
as describing the transfer of particles to a four dimensional
volume element, at $X$, from  a four dimensional volume element,
at $x_1$, caused by the collision processes. If the net transfer
of particles to a point due to collision processes does not
vanish, the momentum at that point changes by the amount equal to
the mass transferred in a unit of time times the longitudinal part
the velocity (the velocity without the rotation part). This is the
meaning of the term on the right hand side of equation
(\ref{colforc}).

In the same way $\hbar\bnabla^y\left.\tilde
D(X,y|x_1)\right|_{y\rightarrow 0}$ stands for the transfer of
momentum from a four dimensional volume element, at $X$, to a four
dimensional volume element, at $x_1$, caused by the collision
processes. It is the quantum mechanical equivalent of the force
applied to part of the system. This is the meaning of the term in
the right hand side of equation (\ref{colforc}). It obeys Newton
third law, $\hbar\bnabla^y\left.\tilde D(X,y|x_1)
\right|_{y\rightarrow 0} = -\hbar\bnabla^y\left.\tilde
D(x_1,y|X)\right|_{y\rightarrow 0}$ only for internal
interactions. If for example an interaction with a phonon bath or
impurities is considered this relation will not hold.

It has long been known that approximations used in transport
theory can in principle lead to transport equations that are not
necessarily consistent with all the macroscopic conservation laws
(see, e.g., Mermin, 1970; Hu \etal , 1989). This possible
inconsistency was sometimes used to estimate the correctness of
the results obtained (Thornber and Feynman, 1970). It is therefore
desired to build the macroscopic conservation laws into the
structure of the approximation used. Papers by Baym and Kadanoff
(1961) and Baym G and Kadanoff (1962) (see also Green \etal ,
1985) found criteria for maintenance of the {\em macroscopic}
conservation laws in systems with electron-electron interaction.
The main motivation for including the conservation laws in the
transport formalism comes from the division of the transport
processes into two types. First, the particles interact via the
interaction specified in the approximation, with this process
generally having a small time constant. Then, the kinetic process
proceeds, with a large time constant, via interactions with an
unspecified bath. This part of the interaction is described by the
{\em macroscopic} conservation laws. Using the hydrodynamic
equations it is possible to build the {\em microscopic}
conservation laws into the structure of the approximation. That
is, equation (\ref{mce2}) holds for every $X$. This is a new
aspect of transport theory that comes out naturally from the
hydrodynamic equations.

As an example, we mention the calculations of the conductivity in
mesoscopic structures based on the quasi-classical Boltzmann
equation. In these calculations the integration over the wave
functions always precedes the evaluation of physical quantities
(e.g., current density). Only after the integration, one equates
the total influence of the electric field to the total influence
of the impurities or of the phonons or of any other interaction.
This procedure leads to {\em macroscopic} momentum conservation in
the system but the {\em microscopic} momentum conservation is
abandoned. In a paper to be soon published, we use the
hydrodynamic equations to look for the current density in
mesoscopic structures keeping the {\em microscopic} momentum
conservation. We found that for elastic impurity s-scattering the
current is close to being constant over the cross-section of the
structure (i.e., the velocity of the electrons is higher near the
edges of the structure). This result differs from the commonly
used results that {\em assume} constant velocity. We deliberately
used the simplest and most widely applied approximations that are
well known and commonly used for decades with the quasi-classical
Boltzmann equation to emphasize the role of the one particular
assumption we changed. We require a {\em microscopic} balance of
forces rather than a {\em macroscopic} balance of forces.

\subsection{A single particle Schr\"odinger equation for a many
particle system}
The hydrodynamic equations were originally derived from the
Schr\"odinger equation without electromagnetic fields by Madelung
(1927) and with electromagnetic fields by Janossy and
Ziegler-Naray (1964). Here we discuss an interesting possibility.
One may introduce a modified Schr\"odinger equation for a wave
function of a single particle excitation in a multi-particle
system which is equivalent to the above hydrodynamic equations. We
mean that it is possible to derive hydrodynamic equations for a
system in electromagnetic fields with interaction, (\ref{mce2}),
from the following modified Schr\"odinger equation,
$$i\h\frac{\partial\Psi(X)}{\partial T}= $$
$$\left( \frac{\left(-i\h\bnabla-(e/c)A(X)+\int^T\di T' {\bf
f}_{coll}^t({\bf X},T')\right)^2}{2m}\right.+e\varphi(X)+ $$
\eqn{ \left. \int^{\bf X}\di {\bf X'}{\bf f}_{coll}^l({\bf X}',T)
+{i\hbar\over\rho(X)}\dint \di \bP I_{coll} (\bP,X)\right)\Psi(X).
}{modif}
Here ${\bf f}_{coll}^t(X)$ and ${\bf f}_{coll}^l(X)$ are the
transverse and the longitudinal components of the collision term
${\bf f}_{coll}(X)$.

Now we demonstrate the derivation of equation (\ref{mce2}) from
equation (\ref{modif}) in two possible ways:
\begin{enumerate}
\item
Introducing  $\rho(X)=\Psi(X)\Psi^*(X)$ and the Hamilton principal
function $S(X)=(-i\h/2)\ln(\Psi(X)/\Psi^*(X))$ (that is writing
$\Psi(X) = \sqrt{\rho(x)} \exp(-i S(X)/\hbar)$) one finds the
continuity equation and the expression for the canonical momentum
$$\dbarbP_{can}(X)=\bnabla\cdot S(X)=$$
\eqn{\left( m\dbarv(X)+(e/c)A(X)+\int^T\di T' {\bf
f}_{coll}^t({\bf X},T') \right) .}{htr}
It should be noted that the velocity is derived from the wave
function by means of the equation
\begin{eqnarray}
\rho (X)\dbarv (X) &=&
\frac{m \dbarJ_{kin} (X)}{e}=
\nonumber \\ &&
{1 \over 2}
\left(-i\h \partial^{X^\prime}-{e\over c}A(X^\prime) -\int^T\di T'
{\bf f}_{coll}^t({\bf X}^\prime,T')
\right.
\nonumber \\ &&
\left.\left. +
i\h \partial^X - {e\over c}A(X) -\int^T\di T'
{\bf f}_{coll}^t({\bf X},T')\right)\left\langle
\Psi^\deg(X')\Psi(X)\right\rangle
\right|_{X'\rightarrow X}
,
\label{curr5}
\end{eqnarray}
In the absence of the interaction terms this equation converges to
the commonly used expression.

Now differentiating equation (\ref{htr}) with respect to time and
using (\ref{modif}) we arrive at the kinetic momentum continuity
equation (\ref{mce2}).
\item The same  Schr\"odinger equation can  be derived in a different
manner. The expectation value of the kinetic energy density
$$\dbarvarep(X)=\frac{-i}{\rho (X)}
\dint \di^4P\varepsilon \rho_W(P,X) $$
satisfies the quantum mechanical equivalent of the classical
Hamilton-Jacobi equation
$$
\dbarvarep(X)=-
\frac{\h^2}{8m\rho(X)} \left( \bnabla^X \right)^2 \rho (X)+
\frac{\dbarbPb (X)}{2m}  +
\dbarvarep_{int}(X)=$$
$$
= \frac{m\dbarv (X)^2 }{2}  +
\frac{\h^2}{8m} \left( \frac{ \bnabla^X {\rho (X)}}
{{\rho (X)}}\right)^2 +
\dbarvarep_{int}(X)
$$
\eqn{
= \frac{m\dbarv (X)^2 }{2}  +
\frac{\h^2}{2m} \frac{ \bnabla^{X2} \sqrt{\rho (X)}}
{\sqrt{\rho (X)}} +
\dbarvarep_{int}(X)
.
}{HJJ}
The last equality is obtained by integrating by parts in order to
show explicitly the role of the quantum potential in changing the
kinetic energy. This step is legitimate since an observable is
obtained after integrating over the spatial coordinates.
Substituting (\ref{htr}) into equation (\ref{HJJ}) the expectation
value of the total energy of the system is expressed as a
functional
$$ \langle E \rangle =\int\di\bX \rho
(X)\left(\dbarvarep (X) +\varphi(X) \right)= $$ \eqn{ \int\di\bX
\Psi^*(X)\left(\frac{\left(-i\h\bnabla-(e/c)A(X) +\int^T\di T'
{\bf f}_{coll}^t({\bf X},T')\right)^2}{2m}
+\varphi(X)+\dbarvarep_{int}(X) \right)\Psi(X), }{hyu}
where $\varphi(X)$ is the electric potential. The Schr\"odinger
equation is obtained by varying equation ({\ref{hyu}}) with
respect to the wave function $\Psi(X)$. This equation is, in fact,
written for the wave function of a one particle excitation in a
many particle system. It contains a complex potential
$\dbarvarep_{int}(X)$ created by all possible scattering
processes. Its imaginary part is simply ${\rm Im}
\dbarvarep_{int}(X) = (\hbar/\rho(X))\dint \di \bP I_{coll}
(\bP,X)$. These sort of terms are discussed, e.g., within the
framework of the optical model (see, e.g., Hodgson, 1963). The
real part, ${\rm Re} \dbarvarep_{int}(X)$ is equal to $\int \di X
f_{coll}^l(X)$ plus linked cluster expansion terms.
\end{enumerate}
\subsection{Rotation of the velocity}
The phase of a single particle wave function represented by the
Hamilton principle function $S(X)$, must be single valued and
satisfy certain boundary conditions. In a collision free system
the boundary conditions lead to the Aharonov-Bohm effect (Aharonov
and Bohm, 1959) and their controversial physical role has been
discussed previously in several papers (Bohm and Hiley, 1979;
W\'{o}dkiewicz, 1984; Liang and Ding, 1987; Peshkin and Tonomura,
1989). It has been argued that the hydrodynamic formulation uses
field strength, rather than potentials, and, hence, it can
describe the Aharonov-Bohm effect only when boundary conditions
are added (e.g., Philippidis \etal , 1981). Here, it will be shown
that the Aharonov-Bohm effect can be derived from the hydrodynamic
equations themselves without a use of potentials. The effect will
be generalized to account for the role of collisions and the
boundary conditions for the velocity in systems with interactions
will be presented.

To find the boundary conditions for the velocity $\dbarv(X)$, its
rotation is calculated. The rotation of equation (\ref{mce2}) and
the identity
\eqn{
{\di\bB\over\di t}={\partial\bB\over\partial t}+
\bnabla\times(\bB\times\bv )+ \bv(\bnabla\cdot\bB)
}{idr}
yield the equation (compare to the derivation of Lorentz force
from the Faraday's law, e.g., by Jackson (1975)
$$
m\frac{\di}{\di T}\oint \dbarv (X)\cdot\di {\bf l}
=\oint \left[-e{\bf E}(X) -{e\over c}\dbarv (X)\times
{\bf B}(X)+{\bf f}_{coll}(X)\right]\cdot\di {\bf l}
$$
\eqn{
={e\over c}\frac{\di}{\di T}\int {\bf B}(X)\cdot\di {\bf s} +
\oint {\bf f}_{coll}(X)\cdot\di {\bf l}.
}{gfr}

Equation (\ref{gfr}) can be represented in the form
$$m\oint \dbarv (X) \cdot\di \bl- m\left.\oint\dbarv (X) \cdot\di
\bl \right|_{\bB\equiv {\bf f}_{coll}\equiv 0} -\int^T \di T \oint
{\bf f}_{coll}(X)\cdot\di {\bf l}=$$
\eqn{  {e\over c}\int\bB(X) \cdot \di \bs }{ojh}
where the integration in the left hand side of the equation is
carried out along the curve bounding the integration surface in
the right hand side.

The term $\left.\oint \left\langle \bv(X) \right\rangle \cdot\di
\bl \right|_{\bB\equiv {\bf f}_{coll}\equiv 0} $ is the rotation
of the velocity for the collisionless system without a magnetic
induction. It can be calculated by means of the continuity
equation and equation (\ref{mce2}) which for a steady $(\di\dbarv
/\di T =0)$, cylindrically symmetric system, coincide with the
Schr\"odinger equation for particles with a well defined angular
momentum. The Bohr-Sommerfeld quantization rule, therefore,
follows
$$ \left.\oint m \dbarv(X)  \cdot\di \bl \right|_{\bB \equiv {\bf
f}_{coll} \equiv 0} = nh$$
with $n$ being an integer. A change by one quantum level in the
rotation of the velocity corresponds to a change of a flux unit
$\Phi_0 = ch/e = c2\pi\hbar /e$.

Equation (\ref{gfr}) can be written with the help of equation
(\ref{htr}) in the form
\eqn{
{1\over\hbar}
\oint\nabla\cdot S(X) = 2\pi n,
}{bgt}
which is, in fact, the condition that the wave function is single
valued.

Equation (\ref{ojh}) is rewritten in the form
\eqn{
m\oint \dbarv (X) \cdot\di \bl
=
nh + {e\over c}\int\bB(X) \cdot \di \bs
+ \int^T \di T \oint {\bf f}_{coll}(X)\cdot\di {\bf l}
.}{oyd}
It states that the sum of the angular momentum of the electron, of
the magnetic induction and of the system, that applies the
collision force, should be quantized. All these three sub-systems
are coupled together and a continuous change in the angular
momentum of one of them results in a continuous change in the
angular momentum of the others. For example, if the magnetic
induction is raised gradually to its finite static value, it
induces an electric field which changes the velocity lines in such
a way as to keep the total angular momentum constant (Lentz law)
(c.f., Peshkin, 1981). Equation (\ref{oyd}) tells us nothing about
sudden changes of the total angular momentum in quanta of $h$.
This is done via coupling to an external system that is not
included in our description (e.g., a current source that drives
the solenoid). The external system that applies a torque to change
the angular momentum of the electrons and the magnetic induction
can change their angular momentum only in quanta of $h$, hence, if
it applies a {\em small} and uneven torque to the magnetic
induction and to the electron, it can change the angular momentum
of the electron but not the total angular momentum. The conserved
angular momentum of the external system in this interaction does
not mean that this is a forceless interaction. The internal
distribution of angular momentum in the external system has
changed.

Although the change of the velocity of the electrons is due to the
induced electric field, {\em the system does not relax to its
state before the fields were applied just as a classical particle
that does not relax to its initial velocity after the field has
been removed.} Relaxation may occur only if a dissipation
mechanism is introduced into the system, via ${\bf f}_{coll}(X)$.
Equation (\ref{gfr}) explicitly shows that the change in
$\dbarv(X)$ can be viewed at as being due to a local action of the
electric field at earlier times, or as a non-local action of the
magnetic field at the present time. The collision force can cause
a non-zero rotation of the velocity in the same manner as it can
be done by an electric field generated by a time dependent
magnetic field.
\section{Examples}
\label{examples}
\subsection{\wqdf\ for a harmonic potential in a permanent
magnetic field\label{Landau}}
\subsubsection{Schr\"odinger equation}
A system with a harmonic potential $m\omega_c^2 z^2/2$ in the $z$
direction and a magnetic induction ${\bf B}=(cm/e)\bomega_H$ in
the $x,y$ plane is considered. Before the application of the
magnetic induction the electron is moving with the momentum ${\bf
p}^0= m\dbarv^0$ in the $x,y$ plane. After its application the
electron velocity changes according to equation (\ref{ojh}) and
becomes
\eqn{\dbarv(z)=\bp^0/m + (e/mc)\bB \times \bz
= \bp^0/m + \bomega_H \times \bz.
}{amd}
The force acting on the electron stems from three sources,
harmonic potential, magnetic induction and the quantum potential.
Putting them all in equation (\ref{mce2}), and assuming that the
system is in a steady state, we find an equation for the balance
of forces in the $z$ direction,
\eqn{ 0=
-m(\bomega_c^2 + \bomega_H^2) z - {\bf p}^0 \times{\bomega }_H
+{\hbar^2 \over 2m}{\partial\over\partial z}\left(
\frac{1}{\sqrt{\rho(z)}}\frac{\partial^2 \sqrt{\rho(z)}}{\partial
z^2} \right). }{polk}
After integration (here, a particular choice of the gauge is made
taken, see equation \ref{bbbbb}) we obtain a time independent
Schr\"{o}dinger equation
\begin{eqnarray}
0&=&{\hbar^2\over 2m} \frac{\partial^2 \sqrt{\rho_n(z)}}{\partial
z^2} - \frac{m({\bomega }_c^2 + {\bomega }_H^2) z^2}{2} -{\bf p}^0
\times{\bomega }_H z + E_n \nonumber
\\
&=& {\hbar^2 \over 2 m}\frac{\partial^2 \sqrt{\rho_n(z)}}{\partial
z^2} - \frac{m({\bomega }_c^2 + {\bomega }_H^2) }{2}\left(
(z+z^0)^2-(z^0)^2 \right) + E_n.
\end{eqnarray}
Here
$$z^0 = \frac{\bp^0 \times{\bomega }_H}
{m \left({\bomega }_H^2+{\bomega }_c^2 \right)}.$$
is the deflection of the center of the effective harmonic
potential due to the Hall effect (it is the point where the Hall
force equals the harmonic force). $E_n$ and $\rho_n(z)$ are the
eigenenergy and the electron density for the $n_th$ eigenstate.

The Hamiltonian can be written in the form
\begin{eqnarray}
\hat H &=& \frac{{\hat \bp}^2-\left(\bp^0 \times \frac{{\bomega
}_H}{\omega_H}\right)^2}{2m} + \frac{m{\bomega }_c^2 \hat z^2}{2}
+ \frac{m{\bomega }_H^2}{2} \left(\hat z + \bp^0
\times\frac{{\bomega }_H}{m \omega_H^2} \right)^2 \nonumber
\\
&=& \frac{ {\hat \bp}^2 - \left(\bp^0 \times \frac{ {\bomega }_H}
{\sqrt{{\bomega }_H^2+{\bomega }_c^2}} \right)^2}{2m} +
\frac{m\left({\bomega }_H^2+{\bomega }_c^2 \right)}{2} \left(\hat
z + z^0 \right)^2 \label{bbbbb}
\end{eqnarray}
If only the harmonic potential is present or if only the magnetic
induction is present, the above Schr\"{o}dinger equation takes a
form similar to that which can be found in many text-books (see,
e.g., Landau and Lifshitz, 1977).

Solutions of this equation read
$$\sqrt{\rho_n(z)} =$$
$$ {1 \over  2^{n/2} (n!)^{1/2}}\left( {m\sqrt{{\bomega }_c^2 +
{\bomega }_H^2}\over\pi\hbar}\right)^{1 \over 4} \exp\left\{
-\frac{m}{2 \hbar}\sqrt{{\bomega }_c^2 + {\bomega }_H^2} \left( z+
z^0 \right)^2 \right\}\times$$
$$ H_n\left(\sqrt{ \frac{m}{\hbar}\sqrt{{\bomega }_c^2 + {\bomega
}_H^2}} \left( z+z^0 \right) \right)$$
where $H_n$ are the Hermite polynomials. The energy levels are
given by
\begin{eqnarray*}
E
&=& (n+1/2)\hbar \sqrt{{\bomega }_c^2 + {\bomega }_H^2}
+ \frac{({\bf p}^0\times{\bomega }_H)^2}
{2m({\bomega }_c^2 + {\bomega }_H^2)}
\\
&=& (n+1/2)\hbar \sqrt{{\bomega }_c^2 + {\bomega }_H^2}
+
\frac{m({\bomega }_c^2 + {\bomega }_H^2)}{2} (z^0)^2.
\end{eqnarray*}
and, as expected, correspond to the levels of the effective
harmonic potential shifted due to the Hall force.

The above derivation emphasizes a connection between the momentum
continuity and the Schr\"{o}dinger equations. An attention should
be also drawn to the velocity profile. It is the velocity (but not
the current) which grows with the coordinate $z$. This can be seen
directly from equation (\ref{amd}) or (more roughly) from the
independence of $\rho(z)$ from the coordinates $x$ and $y$. This
suggests that the canonical momentum is  constant, whereas the
kinetic momentum equals to the vector potential that generates the
magnetic induction.

\subsubsection{\wqdf}
Gauge-invariant \wqdf\ should satisfy the quantum Hamilton-Jacobi
and Boltzmann equations for the system under consideration. The
quantum Hamilton- Jacobi equation is found by substituting the
electric field $\bE=m\omega^2 \bz$ and the magnetic induction
${\bf B}=(cm/e){\bomega }_H$ into equation (\ref{qbehj}),
$${1\over 2\pi\hbar}\left\{ \varepsilon + {\hbar^2 \over 6}
m\omega_c^2 \left(\partial^{p_z}\right)^2 +{\hbar^2 \over 4
m}\left[\bnabla^X - m\omega_c^2 \bz \partial^\varepsilon
+m{\bomega }_H\times\bnabla^p \right]^2 \right. $$
\eqn{ \left. -{1 \over m}\left[\bp- {\bz\over z}{\hbar^2 \over 12}
m\omega_c^2 \partial^{p_z}\partial^\varepsilon\right]^2
\right\}\rho_W(\varepsilon,\bp,\bx)=0. }{dkuy}
Quantum Boltzmann equation (\ref{qbebl}) has the form
$$
\left\{
-m\omega_c^2 \bz \partial^{P_z}
+{1 \over 2m}
\left[\bp- {\bz\over z}{\hbar^2 \over 12} m\omega_c^2
\partial^{p_z}\partial^\varepsilon\right]
\left[\bnabla^X - m\omega_c^2 \bz \partial^\varepsilon +m{\bomega
}_H\times\bnabla^p \right] \right. $$
$$\left. + {1 \over 2m} \left[\bnabla^X - m\omega_c^2 \bz
\partial^\varepsilon +m{\bomega }_H\times\bnabla^p \right]
\left[\bp- {\bz\over z}{\hbar^2 \over 12} m\omega_c^2
\partial^{p_z}\partial^\varepsilon\right]
\right\}\rho_W(\varepsilon,\bp,\bx)$$
\eqn{=0.
}{jbkuy}
The set of functions that solve these equations are
\begin{eqnarray}
\rho_W(\varepsilon,\bp,z|n) &=& \sqrt{{m\sqrt{{\bomega }_c^2 +
{\bomega }_H^2}\over\pi\hbar}} \exp \left[ -{m\sqrt{{\bomega }_c^2
+ {\bomega }_H^2}\over\hbar} \left( (z+Z_0)^2 + \frac{6
\Omega^2}{m\omega_c^2} \right)\right] \nonumber \\ &&
\sqrt{\frac{\hbar^2}{6 m\omega_c^2\Omega^2}} \cos\left({p_z\over
\hbar}\sqrt{\frac{24 \Omega^2}{m\omega_c^2}}\right)
\theta_H(\Omega^2) \nonumber \\ && L_n \left( {4
\over\hbar\sqrt{{\bomega }_c^2 + {\bomega }_H^2}}
\left({m({\bomega }_c^2 + {\bomega }_H^2)^2 \over 2}(z+Z_0)^2 +
\frac{p_z^2}{2 m}\right) \right)\label{jdwa}
\end{eqnarray}
where $L_n$ are the Laguerre polynomials. $\Omega$ is defined by
means of the equation
$$ 0=\Omega^2 + \varepsilon -\left({1 \over 2}+n \right)
\hbar\sqrt{{\bomega }_c^2 + {\bomega }_H^2} + {1
\over2}m\omega_c^2 z^2 -\frac{p_y^2}{2m} -\frac{\left( p_x -
m\omega_H z\right)^2}{2m} \frac{\omega_H}{{\bomega }_c^2 +
{\bomega }_H^2}, $$
and
$$
Z_0=\frac{\left(p_x - m\omega_H z\right)\omega_H}
{m\left({\bomega }_c^2 + {\bomega }_H^2 \right)}.
$$
Integrating equation (\ref{jdwa}) over $\varepsilon$ we find
\begin{eqnarray}
\rho_W(P,z|n)&=&
\exp
\left[
-{2 \over\hbar\sqrt{{\bomega }_c^2 + {\bomega }_H^2}}
\left({m({\bomega }_c^2 + {\bomega }_H^2) \over 2}(z+Z_0)^2
+ \frac{p_z^2}{2 m}
\right)
\right]
\nonumber\\
&&
L_n \left(
{4 \over\hbar\sqrt{{\bomega }_c^2 + {\bomega }_H^2}}
\left({m({\bomega }_c^2 + {\bomega }_H^2) \over 2}(z+Z_0)^2
+ \frac{p_z^2}{2 m}
\right)\right).
\label{gtyu}
\end{eqnarray}

Equation (\ref{gtyu}) is now  compared with other existing
expressions for the \wqdf\ appearing in the literature for a
harmonic oscillator $(\omega_H=0)$ or for a particle in a magnetic
induction $(\omega_c=0)$.

The \wqdf\ and an equation for it in one dimensional harmonic
oscillator has been derived many  times in the past (e.g., Heller,
1976; Bartlett and Moyal, 1949; Groenewald, 1946; Takabayashi,
1954; Carruthers and Zachariasen, 1983). All these derivations
were done for a non-gauge-invariant Wigner function and without
the energy variable. The commonly used equation for the
non-gauge-invariant \wqdf\ for a harmonic oscillator is
$$\left\{ \dbarvarep_{can} + {\hbar^2 \over 8 m} \left[
{\partial^2 \over \partial z^2} + em^2 \omega^2{\partial^2 \over
\partial P^2} \right]  - {P^2 \over 2 m} - {1 \over 2}em\omega^2
z^2 \right\}\rho_W(P,z|n)$$
\eqn{ =0. }{jku}
This equation does not converge to equation (\ref{jdwa})
integrated over $\varepsilon$. However, the solution of
(\ref{gtyu}) satisfies both equation (\ref{jku}) and (\ref{dkuy})
integrated over $\varepsilon$. This situation of one solution for
two different equations, becomes clear if we remember that
$\dbarvarep_{can} = \hbar\omega (n+1/2)$ and that
$$\int\di\varepsilon\;\varepsilon\rho_W(\varepsilon,P,z) = \left[
{7 \over 12}\hbar\omega -{{\bf P}^2 \over 6m} +{1 \over
2}m\omega^2z^2 \right] \int \di \varepsilon \rho_W
(\varepsilon,P,z).$$

We conclude that although the \wqdf, after integrating over
$\varepsilon$, is identical in the gauge-invariant and in the
non-gauge-invariant cases, the equations they obey are different.
This is due to the different expressions for the kinetic and
canonical momenta. Before the  integration over $\varepsilon,$ the
two functions are different as their equations are. The fact that
the two solutions are identical is connected with the fact that
the gauge used in the non-gauge-invariant formalism assumes a
vanishing vector potential, hence, integrating over $\varepsilon$
has washed out all the traces of the gauge.

The \wqdf\ for an ensemble of oscillators at temperature $k_B
T=1/\beta$ can be found using the linearity of the distribution
function.
$$\rho_W(\varepsilon,\bP,\bz|\beta)= \sum_n e^{-(n+{1 \over 2})
\hbar\omega\beta} \rho_W(\varepsilon,\bP,\bz|n)=$$
\begin{eqnarray}
 \sum_n e^{-(n+{1 \over 2}) \hbar\omega\beta} {1 \over\pi\hbar}
\sqrt{{m\sqrt{{\bomega }_c^2 + {\bomega
}_H^2}\over\pi\hbar}}\times\nonumber \\ \exp \left[
-{m\sqrt{{\bomega }_c^2 + {\bomega }_H^2}\over\hbar} \left(
(z+Z_0)^2 + \frac{6 \Omega^2}{m\omega_c^2} \right)\right]\times
\nonumber \\ \sqrt{\frac{\hbar^2}{6 m\omega_c^2\Omega^2}}
\cos\left({P\over \hbar}\sqrt{\frac{24
\Omega^2}{m\omega_c^2}}\right) \times\nonumber\\ L_n \left( {4
\over\hbar\sqrt{{\bomega }_c^2 + {\bomega }_H^2}}
\left({m({\bomega }_c^2 + {\bomega }_H^2)^2 \over 2}(z+Z_0)^2 +
\frac{p_z^2}{2 m}\right) \right) \label{gsje}
\end{eqnarray}
Integrating over $\varepsilon$, using the equality (see, e.g.,
Gradshteyn and Ryzhik, 1980, equation 8.975)
$${1 \over 1-a}e^{{ab\over a-1}} =
\sum_{\imath=0}^\infty L_n(b) a^i
,\;\;\;\;\;\;\;\;\; |a|<1$$
and normalizing, we arrive at an expression that converges to a
well known expression (see, e.g., Hillery \etal , 1984) for
harmonic oscillators (at $\omega_H =0$).
$$ \rho_W(\varepsilon,\bP,\bz|\beta) = \tanh\left(
\frac{\hbar\sqrt{{\bomega }_c^2 + {\bomega }_H^2}\beta}{2}\right)
$$
$$ \exp\left[ -{2 \over\hbar\sqrt{{\bomega }_c^2 + {\bomega
}_H^2}} \tanh\left( \frac{\hbar\sqrt{{\bomega }_c^2 + {\bomega
}_H^2}\beta}{2}\right) \left( {m({\bomega }_c^2 + {\bomega }_H^2)
\over 2}(z+Z_0)^2 + \frac{p_z^2}{2 m} \right) \right]. $$

A comparison for the zero magnetic induction case can be done
after substituting $\omega_c=0$ in equation (\ref{gtyu}).
\begin{eqnarray}
\rho_W(P,z|n)
&=&
\exp
\left[
-{2 \over\hbar {\bomega }_H}
\left({m{\bomega }_H^2 \over 2}(z+Z_0)^2
+ \frac{p_z^2}{2 m}
\right)
\right]
\nonumber\\
&&
L_n \left(
{4 \over\hbar{\bomega}_H}
\left({m{\bomega }_H^2 \over 2}(z+Z_0)^2
+ \frac{p_z^2}{2 m}
\right)\right)
\nonumber\\
&=&   {1 \over\pi\hbar}
\exp
\left[
-{2 \over\hbar {\bomega }_H}
\frac{p_x^2 + p_z^2}{2 m}
\right]
L_n \left(
{4 \over\hbar{\bomega }_H}
\frac{p_x^2 + p_z^2}{2 m}
\right)
\label{gtye}
\end{eqnarray}
where
$$
\varepsilon_{can} = \left({1 \over 2}+n \right)\hbar\omega_H
-\frac{p_y^2}{2m}.
$$
This is the expression used in the literature (see, e.g., book of
Mahan ,1987 \footnote{The `correct' and `transport' Green's
functions used by Mahan are actually non-gauge-invariant and
gauge-invariant forms of the Green's functions.})

\subsection{Wigner quasi-distribution function in a periodic system
\label{KQ}}
It is well known that the concept of a quasi-momentum plays an
important part in the theory of electrons in periodic potentials
(see, e.g., Weinreich, 1965). If the potential is also periodic in
time, then a quasi-energy is to be introduced, Zel'dovich, 1973;
Zel'dovich \etal , 1976). Hence, if a periodic (in space and/or in
time) electromagnetic field is applied to otherwise homogeneous
system, it may be reasonable to use a \wqdf\ formulated in terms
of quasi-momentum and/or quasi-energy. This may provide us a
better physical insight and make calculations easier. Here we
shall try to reformulate the \wqdf\ using the so-called
$kq$-representation, introduced by Zak (1972), assuming
periodicity in space and/or time of the electromagnetic field.
This may shine a new light on the `acceleration theorem' related,
as we shall see, to gauging rather than to electron dynamics.

A $kq$-representation can be constructed for any pair of operators
$\hat A$ and $\hat B$ that satisfy the commutation relation $[\hat
A,\hat B]=-i\h.$ Then a four dimensional constant vector $a_\mu$,
which is the space-time period of our system, allows one to define
the operators
\eqn{
\hat T(a)=\exp (i\hat B^\mu a_\mu) \;\;\;\;\;
{\mbox{ and}}\;\;\;\;\;
\hat\tau (2\pi\h/a) = \exp(i\hat A^\mu 2\pi\h/a_\mu).}
{kq}
Eigen-functions of these operators satisfy the conditions
\eqn{
\hat T(a)\varphi_{kq}(x)= \exp(ik^\mu a_\mu)\varphi_{kq}(x)}
{11}
\eqn{ \hat\tau (2\pi\h/a)\varphi_{kq}(x) = \exp(iq^\mu (2 \pi
\h/a)_\mu) \varphi_{kq}(x) .}{22}
and form a basis of the $kq$-representation.

The operator $\hat A$ is chosen to be the coordinate operator
$\hat x=(\hat t,\hat \br)$, and $\hat B$ to be the energy-momentum
operator $\hat p=(\hat \varepsilon, \hat\bp)$. Then $k$ is the
quasi-energy-momentum and $q$ is the quasi-four-coordinate. Both
the coordinates and momenta, as well as their corresponding
operators, can be written as (Zak, 1972),
\begin{eqnarray}
x =q +(m^\mu a_\mu)a= q+X\nonumber\\
\label{co}\\
p =k +(m^\mu(2\pi\h/a)_\mu)(2\pi\h/a) =k+K\nonumber
\end{eqnarray}
and
\begin{eqnarray}
\hat x = \hat q+\hat X\nonumber\\
\label{op}\\
\hat p =\hat k+\hat K.\nonumber
\end{eqnarray}
Here $m^\mu$ is a vector of integers, $K$ is the momentum
conjugate to the quasi-coordinates and $X$ is the coordinate
conjugate to the quasi-energy-momenta. It is clear that at
$\h/a\rightarrow 0$ the spacing between the discrete values of $K$
vanishes and one approaches the continuous limit, $K\rightarrow p,
q\rightarrow x$. The opposite limit $\h/a\rightarrow \infty$ also
yields $k\rightarrow p$ and $X\rightarrow x$.

The eigenfunction of the operators $\hat\tau (2\pi\h/a)$ and $\hat
T(a)$ can be now represented in the form
$$\varphi_{kq}(x)=$$
\eqn{\left(\frac{V_0}{(2\pi)^4}\right)^{1/2} \sum_{m^\lambda}
\exp\left(ik^\mu a_\mu (m^\lambda a_\lambda)\right)
\delta\left(x_\mu-q_\mu-(m^\lambda a_\lambda)a_\mu\right),}
{eigen}
where $V_0$ is the volume of the four dimensional periodic unit
cell. These eigenfunctions can be used to define the creation and
annihilation operators $\alpha^\deg(k,q),\ \alpha(k,q)$ in the
$kq$-representation in terms of the creation and annihilation
operators $\varphi^\deg(x),\ \varphi(x)$ in the $x$
representation,
\eqn{
\varphi(x)=\int\di k\di q \alpha (k,q)\varphi_{kq}(x).}
{fun}
\eqn{
\alpha (k,q)= \sum_{m^\lambda}\e^{-ik^\mu a_\mu(m^\lambda a_\lambda)}
\varphi\left(q_\mu+(m^\lambda a_\lambda)a_\mu\right)}
{fan}
Transforming the \wqdf\ into the $kq$-representation the matrix
elements
\eqn{
\langle kq\left|p\right|k'q'\rangle =-i\h (\pd/\pd q)
\delta(q'-q)\delta(k'-k)}
{p}
and
\eqn{ \langle kq\left|x\right|k'q'\rangle = \left[i\h (\pd/\pd k)
+ q\right] \delta(q'-q)\delta(k'-k).} {x}
are used. Then the equality
$$\left\langle \Psi^\deg(X+y/2)\Psi(X-y/2) \right\rangle =$$
\eqn{ \int\di k\di q \left\langle \alpha^\deg(k,q) \exp \left(
{i\over\h} y^\mu \partial_\mu^q\right) \alpha(k,q)\right\rangle.
}{dhs}
is obtained and equation (\ref{wqdf1}) in the $kq$-representation
now reads
\eqn{
\rho_W(P,X)=
\int \di y
\e^{-(i/\h)\left({yP}\right)}
\int \di k\di q
\left\langle \alpha^\deg(k,q+y/2)
\alpha(k,q-y/2)\right\rangle,}
{expc5}
as can be verified by substituting equation (\ref{fan}).

One thing that can be learned from equation (\ref{expc5}) is that
if all the particles in the system acquire an additional
quasi-momentum $\delta k$, it will not have any effect on the
\wqdf\ neither will it have an effect on the measurable
quantities. This is follows from the cancellation of the
additional phase factors in equation (\ref{fan}). Hence in order
to cause a change in the physical system, some other factors
should be changed. This statement applies also to the Bloch
representation. To clarify, we give the connection between the
$kq$-representation and the Bloch representation,
\eqn{
\alpha(k,q)=\sum_n B_n(k)\Psi_{nk}(q)
}{kqB}
where the coefficients $B_n(k)$ are the wave functions in the
Bloch representation, and $\Psi_{nk}(q)$ are the Bloch functions.
If a change of the quasi-momentum is necessary which will also
change the momentum one should introduce a function of the
quasi-coordinate and the quasi-momentum, $f(k,q)$ in the
definition of $\alpha(k,q)$. Usually the function
$\exp\int\lambda(k,q)\di t$ is used with $\lambda(k,q)$ chosen to
be the dispersion relation of the electrons with the homogeneous
fields set to zero. The issue of the validity of this procedure is
reviewed by Nenciu (1991), and we shall not dwell on it here. All
that has been said so far, is that a change in the quasi-momentum
cannot by itself lead to a kinetic change in the physical system.
Hence we suggest that a change in the quasi-momentum due to
external field is a gauge effect.

It is sometimes stated, that the dynamics of an electron in a
lattice can be described by the so called acceleration theorem
which is generally written in the form
\eqn{
\frac{\di \bk}{\di t} =-e\bE -(e/c)\bv_{av}\times\bH
}{acc}
where $\bk$ is the quasi-momentum of the particle and $\bv_{av}$
is the average velocity of the electron. It is usually stated that
one may find a basis of eigenfunctions in such a way that it will
make the average velocity of the electron, after a sufficient long
time, to be $\bv_{av}=\bnabla^k\lambda(\bk)$, where $\lambda(\bk)$
is the dispersion relation (Adams and Argyres, 1956; Wilson,
1965).

The acceleration theorem is regarded as a quasi-classical
approximation that can be used only if the external fields are
weak and if the movement of the Bloch electron is restricted to a
small part of the Brillouin zone. The existing derivations of the
acceleration theorem (Wilson, 1965; Landau and Lifshitz, 1984;
Kittel, 1963; Kohn, 1959) are very complicated, controversial and
are not rigorous. Hence, their validity is not well defined. An
exception is the paper of Zak (1972) who has derived the
acceleration theorem for an electric field in a rigorous way.

It is interesting to note a similarity between the acceleration
theorem and equation (\ref{dif}). Here $\bv_{av}$ corresponds to
the velocity of the observer $\di x''/\di t''$ and the canonical
momentum to the quasi-momentum. To make it more clear we remark
that $\dbarbP_{can}(X;x'')$ is equal to the expectation value of
the momentum $\hat p =\hat k+\hat K$. Remembering that the
expectation value of $\hat K$ can take only discrete values, we
realize that there are short periods of time $\Delta t''$, when
the change of the expectation value of the momentum is equal to
the change of the expectation value of $\hat k$. At these periods
of time the acceleration theorem and equation (\ref{dif}) are
identical.
\subsection{Canonical energy change of a heavy mass polaron in
electric field}
The problem of a heavy mass polaron\footnote{The heavy mass
polaron is discussed only for the sake of a demonstration. We
therefore do not provide an introduction to the problem of
polarons and assume that the reader is familiar with the problem.
The background can be found in text-books (see, e.g., Mahan,
1990).} (van Haeringen W, 1965) is solved with the help of the
quantum mechanical equivalent of the classical Hamilton-Jacobi
equation. We shall solve here the problem of a heavy mass polaron
in a constant and homogeneous electric field. This example is also
interesting in and of itself, since this problem has never been
solved by means of a Dyson equation, (see discussion in the book
of Mahan (1990), p.509-510).

We are concerned with a homogeneous system with a single electron
in a conduction band and assume an electron-phonon interaction
with optical phonons. For that kind of system $G^\leq =0$ and only
one Green's function should be evaluated (in the general case
there are two independent Green's functions) and we shall take it
to be $G^c$. This function satisfies the quantum Hamilton-Jacobi
equation (\ref{qbehj}) that for our system takes the form
$${1\over 2\pi\hbar}
\left[ \varepsilon - \varepsilon_{\bf p} +\frac{\hbar^2}{8m}
(\bnabla^\bX + e{\bf E}\partial^\varepsilon)^2\right] G^c({\bf
p},\varepsilon,\bX)=$$
\eqn{ 1 + \Sigma^c({\bf p},\varepsilon,\bX)G^c({\bf
p},\varepsilon,\bX) }{pol}
with $\varepsilon_{\bf p}= {\bf p}^2/(2m)$. The right hand side of
the equation takes this simple form since the polaron has a heavy
mass and all the derivatives of the Green's function with respect
to the momenta are assumed to vanish. Hence, the phase loop
function (that represents kinetic effects) vanishes. The
self-energy term has the general form
\eqn{ \Sigma^c({\bf p},\varepsilon) = \int\di \omega \di {\bf q}
G^c({\bf p},\varepsilon-\omega - \varepsilon_{\bf q})D^c({\bf
q},\omega) \Gamma({\bf p},\varepsilon,{\bf q},\omega) }{self}
$D^c({\bf q},\omega)$ is the phonon Green's function and the
vertex part is given by the Ward identity (Engelsberg and
Schrieffer, 1963)
\eqn{
\Gamma({\bf p},\varepsilon,{\bf q},\omega)=
\frac{
\left[ {G^c}({\bf p},\varepsilon-\omega - \varepsilon_{\bf q})
\right]^{-1}
-
\left[ {G^c}({\bf p},\varepsilon)
\right]^{-1}
}{
\omega + \varepsilon_{\bf q}
}
}{ward}

Fourier transforming equation (\ref{pol}), one obtains
\eqn{
\left[-i\hbar{\di\over \di t} - \varepsilon_{\bf p}
+\frac{1}{8m}\left(-i\hbar\bnabla^\bX+e{\bf E} t\right)^2
-\Phi(t)\right]G^c({\bf p},t)= \hbar\delta (t).
}{tt}
where
$$ \Phi(t)= \int^t \di T \di {\bf q}D^c({\bf q},T)
e^{i\varepsilon_{\bf q}T}. $$
The solution of this equation is
\eqn{
G^c({\bf p},t,\bX)=
\theta(t)\exp\left[{i\over \hbar} \left(
\left(\varepsilon_{\bf p}+e\bE\bX\right) t
+ \int^t \di T \Phi(T)\right)\right]
.}{pol1}
As can be seen from this equation the energy of the polaron is
composed of three parts corresponding to its kinetic, potential
and phonon interaction energies. It is instructive to search for a
solution of equation (\ref{pol}) that does not depend on $\bX$
\eqn{ {1\over 2\pi\hbar} \left[ \varepsilon - \varepsilon_{\bf p}
+ \frac{\hbar^2}{8m} (e{\bf E}\partial^\varepsilon)^2\right]
G^c({\bf p},\varepsilon)= 1 + \Sigma^c({\bf
p},\varepsilon)G^c({\bf p},\varepsilon). }{pol9}
It reads
\eqn{G^c({\bf p},t)= \theta(t)\exp\left[{i\over \hbar}
\left(\varepsilon_{\bf p} t- \frac{(e{\bf E})^2}{24m} t^3 + \int^t
\di T \Phi(T)\right)\right] }{pol19}
Going back to equation (\ref{dif}) we see that the change of the
canonical momentum of the particle at rest in a static and
homogeneous electric field is $e{\bf E}t$. The change of the
canonical energy due to the electric field is
$$\varepsilon_{can}(t)= \int^t \di t_1 {e\over 2m}{\bf E} {\bf
P}(t_1)= {1\over m}\left({e\over 2}{\bf E}\right)^2\int^t \di t_1
t_1= {1\over m}\left({e\over 2}{\bf E}\right)^2{t^2\over 2}.$$
Integrating over time $t$ we arrive at the phase factor in
equation (\ref{pol1}) which can be now written in the form
\eqn{ G^c({\bf p},t)= \theta(t)\exp\left[{i\over \hbar}
\left(\int^t \left( \varepsilon_{\bf p} - \varepsilon_{can}(T)+
\Phi(T)\right)\di T\right)\right].}{pol2}
It does not contain the potential energy term. There is a new term
instead which corresponds to the energy that a free particle would
acquire if it were moving freely in the electric field.

\section{Summary}
In this paper we formulate a gauge-invariant formalism for
transport processes starting from the definition of the
gauge-invariant \wqdf, we also derive quantum transport equations
and their hydrodynamic representation. Throughout the paper we
emphasized the kinetic vs canonical meaning of the quantities
discussed. We believe that this formulation will both simplify
studying transport phenomena and reveal new aspects that are not
easily shown with other available formalisms (e.g.,
quasi-classical Boltzmann equation or Kubo correlators).

\vspace{1cm}

\appendix{\bf APPENDIX A}

\vspace{1cm}

\setcounter{equation}{0}
\renewcommand{\theequation}{A.\arabic{equation}}
The equality
$$\exp\left({-{i\over\h} \hat P_{can}^\mu \hat y_\mu}\right)
\exp\left({{i\over\h} \hat P_{kin}^\mu \hat y_\mu}\right) =$$
\eqn{ \exp\left( {i e\over c\h}\dint^{1}_{0} \hat y^\mu A_\mu
 (\hat X+s\hat y)\di s \right)}{eqll1}
in which the canonical and kinetic momenta are connected by the
equation
\eqn{
\hat P_{kin}= \hat P_{can}-{e\over c} A(\hat X)
=-i\h{\pd\over\pd X} -{e\over c}A(\hat X).}
{mom}
is derived.

First, the operator
\eqn{
U(s)=
\exp\left({-{i\over\h}s \hat P_{can}^\mu \hat y_\mu}\right)
\exp\left({{i\over\h}s \hat P_{kin}^\mu \hat y_\mu}\right)}
{u}
is defined. It satisfies the following two conditions
\eqn{
\left. U(s)\right|_{s=0}=1}
{der}
and
\eqn{ \frac{\pd U(s)}{\pd s} = {i e\over c\h} \hat y^\mu
A_\mu(\hat X+s\hat y) U(s)} {dft}
where
\eqn{ A(\hat X+s\hat y) = \exp\left({-{i\over\h}s\hat
P_{can}^\mu\hat y_\mu}\right) A(\hat X) \exp\left({{i\over\h}s\hat
P_{can}^\mu\hat y_\mu}\right).} {aA}
Then integrating equation (\ref{dft}) the equality (\ref{eqll1})
is obtained.

\vspace{1cm}

\appendix{\bf APPENDIX B}

\vspace{1cm}

\setcounter{equation}{0}
\renewcommand{\theequation}{B.\arabic{equation}}
The phase loop function was introduced in our previous paper
(Levanda and Fleurov, 1994). Here we consider its shape for
systems in a static and homogeneous field. The phase loop function
is another representation of the slashed derivatives appearing in
equations (\ref{qbehj}) and (\ref{qbebl}) and discussed in detail
in main body of the paper.

The phase loop function corresponding to an electron loop with
three vertices is defined by the equation
\begin{eqnarray}
\lefteqn{
M({\bpi}\mid 3,F(X))
= M(\{\pi_i\}\mid F(X)) =}
\nonumber \\
& &
\dint\di^4x_1\di^4x_2\di^4x_3
\exp\left({i\over\h}\sum_{i=1}^3\pi_i^\mu x_{i\mu}\right)\cdot
\exp\left({ie\over c\h}\Phi\left(\{x_i\}\right)\right)
\label{Green4}
\end{eqnarray}
\marginpar{Green4}
where
\eqn{
\Phi\left(\{x_k\}\right)=
\sum^3_{i=1}y_k^\mu \dint_{-{1/2}}^{1/2}
A_\mu (X_k+s_k y_k) \di s_k ,}
{Phi}
$y_k= x_{k+1} - x_k$, $X_k = (x_{k+1}+x_k)/2$, and cyclic
conditions $x_{3+i} = x_i$ being assumed. Such a phase loop
function depends on three four-dimensional variables
$\{\pi_i^\mu\}^3_{i=1}$, on the electromagnetic field tensor
$F^{\lambda\mu}(X)$ and is explicitly gauge invariant. In the
absence of electromagnetic fields the phase loop function takes
its trivial form
\eqn{ M({\bpi}\mid 3,F(X)) =\prod_{\mu,i}\delta(\pi^\mu_i).}
{triv}
Choosing  $x''=x_1$ in equation (\ref{int22}) we find that
\eqn{
\Phi\left(\{x_k\}\right)=
{1\over 4}{\cal F}^{\lambda\mu}\left[
(x_2-x_1)_\mu (x_3-x_1)_\lambda -
(x_3-x_1)_\mu (x_2-x_1)_\lambda
\right].}
{Phi3}

This is just an expression for the flux through a triangle that
has its three vertices at the points $x_1,x_2,x_3$,
\eqn{
\Phi\left(\{x_k\}\right)=
\dint_{S\{x_1,x_2,x_3\}}
F^{\lambda\mu}\di x_\mu \wedge \di x_\lambda, }
{Phi2}
where
$S\{x_1,x_2,x_3\}$
is the surface covering this triangle.

The integration in the definition of the phase loop function is
now straightforward and results in
$$M({\bpi}\mid 3,{\cal F_{\mu\lambda}})=$$
\eqn{
\delta\left(\sum_{i=1}^3\bpi_{i}/\h\right)
\exp\left[
(-ie\h/4c){\cal F_{\mu\lambda}}
\pd^{\pi_2\mu}\pd^{\pi_3\lambda}
\right]
\prod_{i=2}^3\delta(\bpi_{i}/\h).}
{notri}
The latter equation (\ref{notri}) can be easily generalized to a
$N$ vertex electron loop in a homogeneous static  field. The
polygon with $N$ vertices can be subdivided into $N-2$ triangles
and the phase loop function for such $N$ vertex electron loop is
$$M(\bpi\mid N,{\cal F_{\mu\lambda}})=$$
\eqn{
\delta\left(\sum_{i=1}^N\bpi_{i}/\h\right)
\exp\left[
-i(e\h/4c){\cal F_{\mu\lambda}}
\sum_{k=2}^{N-1}
\pd^{\pi_k\mu}\pd^{\pi_{k+1}\lambda}
\right]
\prod_{i=2}^N\delta(\bpi_{i}/\h).}
{notriv}

Comparing this equation with equation (\ref{triv}), one realizes
that the momentum or energy are no longer conserved at each
vertex. In this representation, the phase loop function shows in a
simple way, how the field interferes with the four-momentum
conservation. Derivatives of the delta function mean that
derivatives of the Green's functions and of the vertex functions
appear in analytical expressions for the diagrams. These
derivatives stand for the electron recoil due to the field
(compare this to the recoil expansion of van Hearingen, 1965).
Equation (\ref{notriv}) converges to equation (\ref{triv}) as
${\cal F_{\mu\lambda}}$ approaches zero.

{\Large References}

Adams E N and P N Argyres 1956 {\em Phys. Rev.} {\bf 102} 605

Aharonov Y and D Bohm 1959 {\em Phys. Rev.} {\bf 115} 485

Al'tshuler B L 1978 {\em Sov. Phys. JETP} {\bf 48} 670

Balazs N L and B K Jennings 1984 {\em Phys. Rep.} {\bf 104} 347

Bartlett M S and J E Moyal 1949 {\em Proc. Camb. Phil. Soc. math.
phys. Sci.} {\bf 45} 545

Baym G 1962 {\em Phys. Rev.} {\bf 127} 1391

Baym G and L P Kadanoff 1961 {\em Phys. Rev.} {\bf 124} 287

Bertoncini R, A M Kriman and D K Ferry 1989 {\em Phys. Rev. B}
{\bf 40} 3371

Berry M V 1977 {\em Phil. Tran. R. Soc. London} {\bf 287} 30

Bohm D 1952 {\em Phys. Rev.} {\bf 85} 166, 180

Bohm D and Hiley B J 1979 {\em Nuovo Cimento A} {\bf 52} 295

Bund G W 1995 {\em J. Phys. A: Math. Gen.} {\bf 20} 3709

Carruthers P and F Zachariasen 1983 {\em Rev. Mod. Phys.} {\bf 55}
245

Cohen L 1966 {\em J. Math. Phys. }{\bf 7} 781

Cohen L 1976 {\em J. Math. Phys. }{\bf 17} 1863

Cohen L 1987 in {\em The Physics of Phase Space } ed. Y S Kim and
W W Zachary Springer- Verlag: Berlin

Davies J H and J W Wilkins 1988 {\em Phys. Rev. B} {\bf 38} 1667

Elze H-Th, M Gyulassy and D Vasak 1986 {\em Nucl. Phys. B} {\bf
276} 706

Engelsberg S and J R Schrieffer 1963 {\em Phys.Rev.} {\bf 131} 993

Fleurov V N and A N Kozlov 1978 {\em J. Phys. F: Metal Phys} {\bf
8} 1899

Ferry D K and H L Grubin 1995 in {\em Solid State Physics} {\bf
45} ed: H Ehreriech and F Spaepen (Academic Press:N.Y.)

Gradshteyn I S and I M Ryzhik 1980 {\em Table of Integrals, Series
and Products} (Academic Press: N.Y.)

Green F, D Neilson and J Szymanski 1985 {\em Phys. Rev. B} {\bf
31} 2779

Groenewald H J 1946 {\em Physica} {\bf 12} 405

van Haeringen W 1965 {\em Phys. Rev.\ }{\bf 137} A 1902

Heller E J 1976 {\em J. Chem. Phys.} {\bf 65} 1289

Heller E J 1977 {\em J. Chem. Phys.} {\bf 67} 3339

Hillery M, R F O'Connell, M O Scully and E P Wigner 1984 {\em
Phys. Rep.} $\hspace{8em}$ {\bf 106} 121

Hodgson P E 1963 {\em The optical model of elastic scattering}
(Clarendon: Oxford)

Holland P R 1993 {\em The Quantum Theory of Motion} Cambridge
Univ. Press

Hu B Y K, S K Sarker, J W Wilkins 1989 {\em Phys. Rev. B} {\bf 39}
8468

Irving J H 1965 {\em Wigner distribution function in relativistic
quantum mechanics } Ph.D. thesis, Princeton Univ.

Jackson J D 1975 {\em Classical Electromagnetics} Second Edition,
Wiley

Janossy L and M Ziegler-Naray 1964 {\em Acta Phys. Hung.} {\bf
XVI} 345

Kadanoff L P and Baym G 1962 {\em Quantum Statistical Mechanics}
(New York: Benjamin)

Khan F S, J H Davies and J W Wilkins 1987 {\em Phys. Rev. B} {\bf
36} 2578

Keldysh L V 1965 {\em Sov. Phys. JETP} {\bf 20} 1018

Kittel C 1963 {\em Quantum Theory of Solids} Wiley

Kohn W 1959 {\em Phys. Rev. }{\bf 115} 1460

Kubo R 1957 {\em J. Phys. Soc. Japan} {\bf 12} 570

Landau L D and E M Lifshitz 1977 {\em Quantum Mechanics: Non
Relativistic Theory} (Pergamon Press)

Landau L D and E M Lifshitz 1984 {\em Statistical Mechanics vol 2}
(Pergamon Press)

Lee H-W and M O Scully 1983 {\em Foundations of Physics} {\bf 13}
61

Levanda M and V Fleurov 1994 {\em J. Phys.: Condens. Matter} {\bf
6} 7889

Liang J Q and X Ding 1987 {\em Phys. Lett. A} {\bf 119} 325

Mahan G D 1987 {\em Phys. Rep.} {\bf 145} 251

Mahan G D 1990 {\em Many Particle Physics} (New York: Plenum
Press)

Madelung E 1927 {\em Z. Phys.} {\bf 40} 322

Mandelstam S 1960 {Ann. Phys. (N.Y.)} {\bf 19} 1

Martin P C and J Schwinger 1959 {\em Phys. Rev.} {\bf 115} 1342

McCoy N H 1932 {\em Proc. Nat. Acad. Sci.} {\bf 18} 674

Mermin N D 1970 {\em Phys. Rev. B} {\bf 1} 2362

Messiah A 1962 {\em Quantum Mechanics} North-Holland

Moyal J E 1949 {\em Proc. Cambridge Phil. Soc.} {\bf 45} 99

Nenciu G 1991 {\em Rev. Mod. Phys.} {\bf 63} 91

Peshkin M 1981 {\em Phys. Rep.} {\bf 80} 375

Peshkin M and A Tonomura 1989 {\em The Aharonov -- Bohm Effect},
Lecture Notes in Physics, V.340, Springer Verlag, Berlin

Philippidis C, D Bohm and R D Kaye 1981 {\em Il Nuovo Cimento}{\bf
71}75

Pimpale A and M Razavy 1988 {\em Phys. Rev. A }{\bf 38} 6046

Rammer J and H Smith 1986 {\em Rev. Mod. Phys.} {\bf 58} 323

Reggiani L, P Lugli and A P Jauho 1987 {\em Phys. Rev.} {\bB} {\bf
36} 6602

Reizer M Yu and A V Sergeev 1987 {\em Sov. Phys. JETP}\ {\bf 66}
1250

Rohrlich F 1965 {\em Classical Charged Particle} Addison- Wesley:
Massachusetts

Scully M O and L Cohen 1987 in {\em The Physics of Phase Space}
ed. Y S Kim and W W Zachary Springer- Verlag: Berlin

Schwinger J 1951 {\em Phys. Rev.} {\bf 82} 664

Schwinger J 1960 {\em Proc. Nat. Acad. Sci.} {\bf 46} 570, 883,
1401

Serimaa O T, J Javanainen and S Varo 1986 {\em Phys. Rev. A} {\bf
33} 2913

Summerfield G C and P F Zweifel 1969 {\em J. Math. Phys. }{\bf 10}
233

Takabayashi T 1954 {\em Prog. Theor. Pys.} {\bf 11} 341

Thornber K K and R P Feynman 1970 {\em Phys. Rev. B} {\bf 1} 4099

Tugushev V V and V N Fleurov 1983 {\em Sov. Phys. JETP} {\bf 57}
1322

Vasak D, M Gyulassy and H-Th Elze 1987 {\em Ann. Phys.}(N.Y.) {\bf
173} 462

van Vleck J H 1928 {\em Proc. N.A.S.} {\bf 14} 178

Weinreich G 1965 {\em Solids: Elementary Theory for Advanced
Students} (John-Wiley: New York)

Weyl H 1931 {\em The Theory of Groups and Quantum Mechanics}
English translation, Dover Publ. Inc.

Wigner E P 1932 {\em Phys. Rev. } {\bf 40} 749

Wilson A H 1965 {\em The Theory of Metals} (Cambridge Univ.
Press.: London)

W\'{o}dkiewicz K 1984 {\em Phys. Rev. A} {\bf 29} 1527

Zak J 1972 in {\em Solid State Physics} {\bf 27,} 1

Zel'dovich Ya B 1973 {\em Soviet Phys. Usp. }{\bf 16} 427

Zel'dovich Ya B, N L Manakov and L P Rapoport 1976 {\em Soviet
Phys. Usp. }{\bf 18} 920

Ziman J M 1960 {\em Electrons and Phonons} (Clarendon: Oxford)

\end{document}